\documentclass[10pt,twocolumn]{IEEEtran}
\usepackage{latexsym}       
\usepackage{graphicx}       
\usepackage{subfigure}
\usepackage{epsfig}
\usepackage{times}
\usepackage{amssymb}
\usepackage{amsbsy}
\usepackage{amsmath}
\usepackage{epsf}

\newcommand{\gb}{\beta}
\newcommand{\grg}{\gamma}

\newcommand{\gl}{\lambda}

\newcommand{\gn}{\nu}

\newcommand{\gr}{\rho}
\newcommand{\grh}{\mbox{$ \hat\rho$}}
\newcommand{\gs}{\sigma}

\newcommand{\gt}{\tau}

\def\bm#1{\mbox{\boldmath $#1$}}
\newcommand{\vga}{\mbox{$\bm \alpha$}}

\newcommand{\vgx}{\mbox{$\bm \xi$}}

\newcommand{\mgG}{\mbox{$\bm \Gamma$}}

\newcommand{\rH}{^{ \raisebox{1pt}{$\rm \scriptscriptstyle H$}}}

\newtheorem{theorem}{Theorem}[section]
\newtheorem{lemma}{Lemma}[section]

\newtheorem{prop}{Proposition}[section]
\newtheorem{claim}{Claim}[section]

\newtheorem{definition}{Definition}[section]

\newtheorem{question}{Question}[section]
\newtheorem{coro}{Corollary}[section]

\newcommand{\beq}{\begin{equation}}
\newcommand{\eeq}{\end{equation}}

\newcommand{\bea}{\begin{array}}
\newcommand{\ena}{\end{array}}
\newcommand{\bds}{\begin {itemize}}
\newcommand{\eds}{\end {itemize}}
\newcommand{\bdf}{\begin{definition}}
\newcommand{\blm}{\begin{lemma}}
\newcommand{\edf}{\end{definition}}
\newcommand{\elm}{\end{lemma}}
\newcommand{\bthm}{\begin{theorem}}
\newcommand{\ethm}{\end{theorem}}
\newcommand{\bprp}{\begin{prop}}
\newcommand{\eprp}{\end{prop}}
\newcommand{\bcl}{\begin{claim}}
\newcommand{\ecl}{\end{claim}}
\newcommand{\bcr}{\begin{coro}}
\newcommand{\ecr}{\end{coro}}
\newcommand{\bquest}{\begin{question}}
\newcommand{\equest}{\end{question}}

\newcommand{\larrow}{{\larrow}}

\def\Re{\ensuremath{\hbox{Re}}}
\def\Im{\ensuremath{\hbox{Im}}}

\newcommand{\diag}{\ensuremath{\mathrm{diag}}}

\newcommand{\argmin}{\ensuremath{\mathrm{arg}\min}}
\newcommand{\argmax}{\ensuremath{\mathrm{arg}\max}}

\newcommand{\Vhat}{{\ensuremath{{\hat{V}}}}}

\newcommand{\va}{{\ensuremath{\mathbf{a}}}}
\newcommand{\vab}{{\ensuremath{\mathbf{\bar{a}}}}}
\newcommand{\vah}{{\ensuremath{\mathbf{\hat{a}}}}}

\newcommand{\vq}{{\ensuremath{\mathbf{q}}}}
\newcommand{\vr}{{\ensuremath{\mathbf{r}}}}
\newcommand{\vs}{{\ensuremath{\mathbf{s}}}}

\newcommand{\vt}{{\ensuremath{\mathbf{t}}}}

\newcommand{\vw}{{\ensuremath{\mathbf{w}}}}
\newcommand{\vwh}{{\ensuremath{\mathbf{\hat{w}}}}}

\newcommand{\vzero}{\ensuremath{\mathbf{0}}}

\newcommand{\bs}{{\ensuremath{\mathbf{s}}}}

\newcommand{\bw}{{\ensuremath{\mathbf{w}}}}

\newcommand{\mA}{{\ensuremath{\mathbf{A}}}}

\newcommand{\mB}{{\ensuremath{\mathbf{B}}}}

\newcommand{\mC}{{\ensuremath{\mathbf{C}}}}

\newcommand{\mF}{{\ensuremath{\mathbf{F}}}}

\newcommand{\mI}{{\ensuremath{\mathbf{I}}}}

\newcommand{\mR}{{\ensuremath{\mathbf{R}}}}

\newcommand{\mRh}{{\ensuremath{\mathbf{\hat{R}}}}}

\newcommand{\mV}{{\ensuremath{\mathbf{V}}}}

\def\IC{\mathbb C}
\def\IN{\mathbb N}
\def\IZ{\mathbb Z}
\def\IR{\mathbb R}

\def\shat{^{\mathchoice{}{}%
 {\,\,\smash{\hbox{\lower4pt\hbox{$\widehat{\null}$}}}}%
 {\,\smash{\hbox{\lower3pt\hbox{$\hat{\null}$}}}}}}

\def\bSigma{{
      \ooalign{
      \smash{\hskip.4pt\raise.4pt\hbox{$\Sigma$}}\vphantom{}\crcr
      \smash{\hskip.7pt\raise.6pt\hbox{$\Sigma$}}\vphantom{}\crcr
      \smash{\hbox{$\Sigma$}}\vphantom{$\Sigma$}}
      \vphantom{\hbox{$\Sigma$}}
      }}
\def\bTheta{{
      \ooalign{
      \smash{\hskip.5pt\raise.5pt\hbox{$\Theta$}}\vphantom{}\crcr
      \smash{\hskip.0pt\raise.1pt\hbox{$\Theta$}}\vphantom{}\crcr
      \smash{\hbox{$\Theta$}}\vphantom{$\Theta$}}
      \vphantom{\hbox{$\Theta$}}
      }}
\def\bDelta{{
      \ooalign{
      \smash{\hskip.4pt\raise.4pt\hbox{$\Delta$}}\vphantom{}\crcr
      \smash{\hskip.7pt\raise.6pt\hbox{$\Delta$}}\vphantom{}\crcr
      \smash{\hbox{$\Delta$}}\vphantom{$\Delta$}}
      \vphantom{\hbox{$\Delta$}}
      }}

\makeatletter

\def\bordermatrix#1{\begingroup \m@th
  \@tempdima 8.75\p@
  \setbox\z@\vbox{%
    \def\cr{\crcr\noalign{\kern2\p@\global\let\cr\endline}}%
    \ialign{$##$\hfil\kern2\p@\kern\@tempdima&\thinspace\hfil$##$\hfil
      &&\quad\hfil$##$\hfil\crcr
      \omit\strut\hfil\crcr\noalign{\kern-\baselineskip}%
      #1\crcr\omit\strut\cr}}%
  \setbox\tw@\vbox{\unvcopy\z@\global\setbox\@ne\lastbox}%
  \setbox\tw@\hbox{\unhbox\@ne\unskip\global\setbox\@ne\lastbox}%
  \setbox\tw@\hbox{$\kern\wd\@ne\kern-\@tempdima\left[\kern-\wd\@ne
    \global\setbox\@ne\vbox{\box\@ne\kern2\p@}%
    \vcenter{\kern-\ht\@ne\unvbox\z@\kern-\baselineskip}\,\right]$}%
  \null\;\vbox{\kern\ht\@ne\box\tw@}\endgroup}
\makeatother

\newcommand{\DL}{\begin{dashlist}}
\newcommand{\DLE}{\end{dashlist}}

\makeatletter
\def\argmin{\mathop{\operator@font arg\,min}}
\def\argmax{\mathop{\operator@font arg\,max}}
\makeatother

\newcommand{\SI}{\begin{indlist}}
\newcommand{\EI}{\end{indlist}}

\begin{document}
\title{Image formation in synthetic aperture radio telescopes}
\author{Ronny Levanda$^a$ and Amir Leshem$^{a,b}$
\thanks{$^a$ School of Engineering, Bar-Ilan University, 52900, Ramat-Gan, Israel. \ \ \
$^b$ Faculty of EEMCS, Delft University of Technology, Delft, The Netherlands.
 Email: leshema@eng.biu.ac.il . Amir Leshem was partially supported by NWO-STW grants 10459, and DTC.5893 (VICI-SPCOM).}}

\maketitle
\begin{abstract}
Next generation radio telescopes will be much larger, more
sensitive, have much larger observation bandwidth and will be
capable of pointing multiple beams simultaneously.
Obtaining the sensitivity, resolution and dynamic range supported by the receivers requires
the development of new signal processing techniques for array and atmospheric calibration as well
as new imaging techniques that are both more accurate and computationally efficient since
data volumes will be much larger. This paper provides a tutorial overview of existing image formation
techniques and outlines some of the future directions needed for information extraction from future radio telescopes.
We describe the imaging process from measurement equation until deconvolution, both as a Fourier inversion problem and
as an array processing estimation problem. The latter formulation enables the development of more advanced techniques
based on state of the art array processing. We demonstrate the techniques on simulated and measured radio telescope data.
\end{abstract}
\section{Introduction}
\begin{figure}
\center{
 \includegraphics[width=0.35\textwidth]{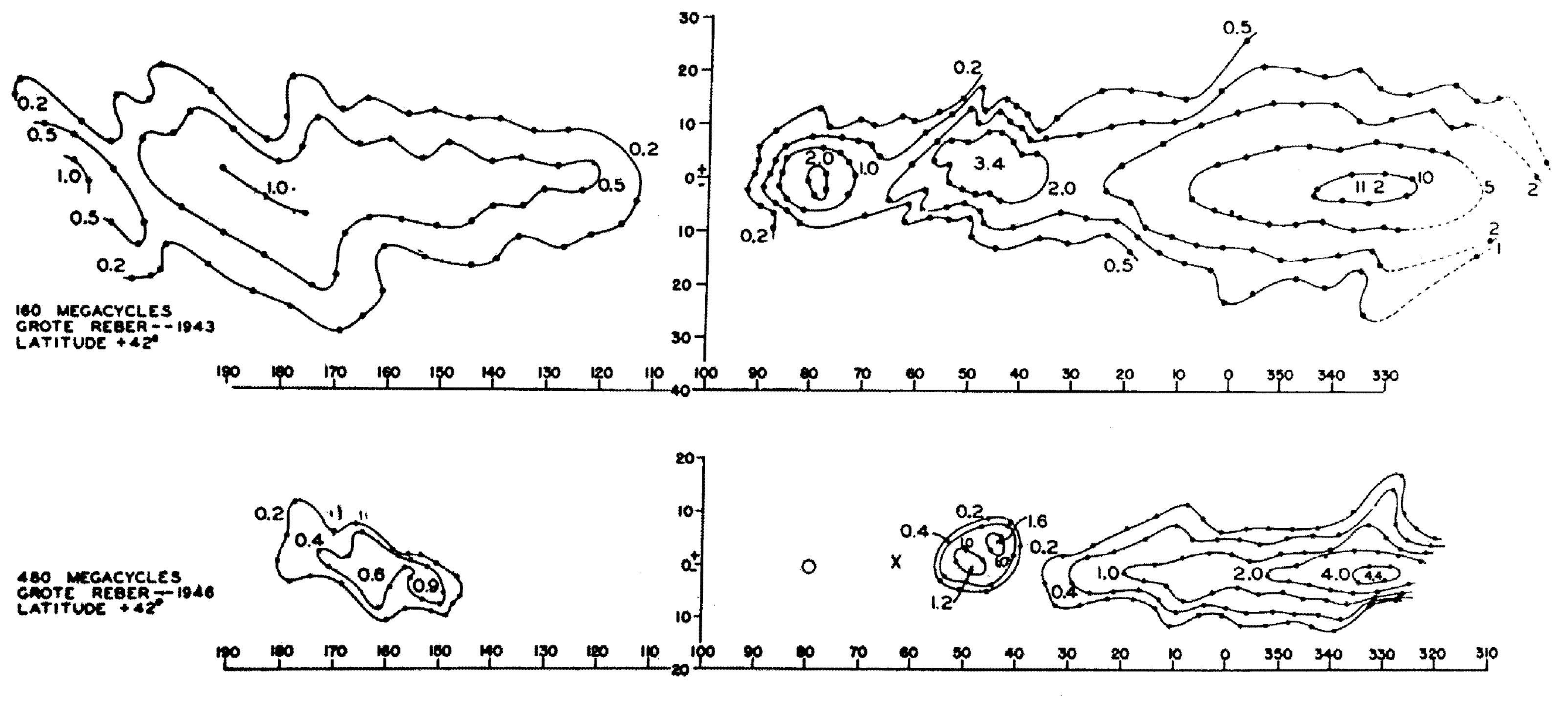}
  \caption{Reber's radio survey. We can see the Milky way Galaxy, Cygnus A and Cassiopeia A. Image courtesy of NRAO/AUI }
 \label{reber_survey}
}
\end{figure}

The field of radio astronomy is a relatively young field of
observational astronomy, dating back to pioneering research by
Jansky in the 1930s \cite{jansky33} who
demonstrated that radio waves are emitted from the Milky Way
galaxy. Inspired by his work, Reber \cite{reber1940} made the first
radio survey of the sky using a radio telescope that he built in his
backyard. Figure \ref{reber_survey} depicts some results of his radio survey, including the strong
radio emissions of Cygnus A (Cyg A) and Cassiopeia A (Cas A).
In 1946 Ryle and Vonberg \cite{ryle1946} used the
Michelson interferometer to observe radio emissions from the sun at
a frequency of 175 MHz. Ryle continued to construct interferometers located on rails, which allowed him
to create a synthetic aperture by moving the antennas. This is the origin of modern inverse synthetic aperture
radar (ISAR) and the active synthetic aperture radar (SAR) imaging.
Subsequently, the study of radio emissions from celestial sources has led to many great discoveries such as
cosmic microwave background radiation by Penzias and Wilson \cite{penzias1965}
and its anisotropy \cite{smoot1992} and pulsars, which are rapidly rotating neutron stars, by Bell et al.
\cite{hewish1968}. Other phenomena of great interest for
radio astronomers include gravitational lenses where the gravitational field of a massive object
serves as a lens by bending the light wave (many of the gravitational lenses were
discovered in radio frequencies \footnote{see
http://www.aoc.nrao.edu/~smyers/class.html}), active galactic nuclei
such as in Virgo A (also known as M87)\footnote{Virgo A is a giant galaxy in the Virgo
cluster which has jets of particles moving at relativistic speeds
and emitting very strong radio waves. It is believed that the center of the Virgo A galaxy is a
very massive black hole} , and supernova remnants
such as Cassiopeia A. Radio astronomy also deals with
spectral lines that appear at radio frequencies such as the Hydrogen
spectral line which was first detected in 1951 \cite{ewen1951}
\footnote{The spectral line at 21 cm is created by a change in the energy state of neutral hydrogen}.
 This spectral line is expected to play an important role in understanding the reionization of the universe
 when the first galaxies were formed.
\begin{figure}
\center{
 \includegraphics[width=0.35\textwidth]{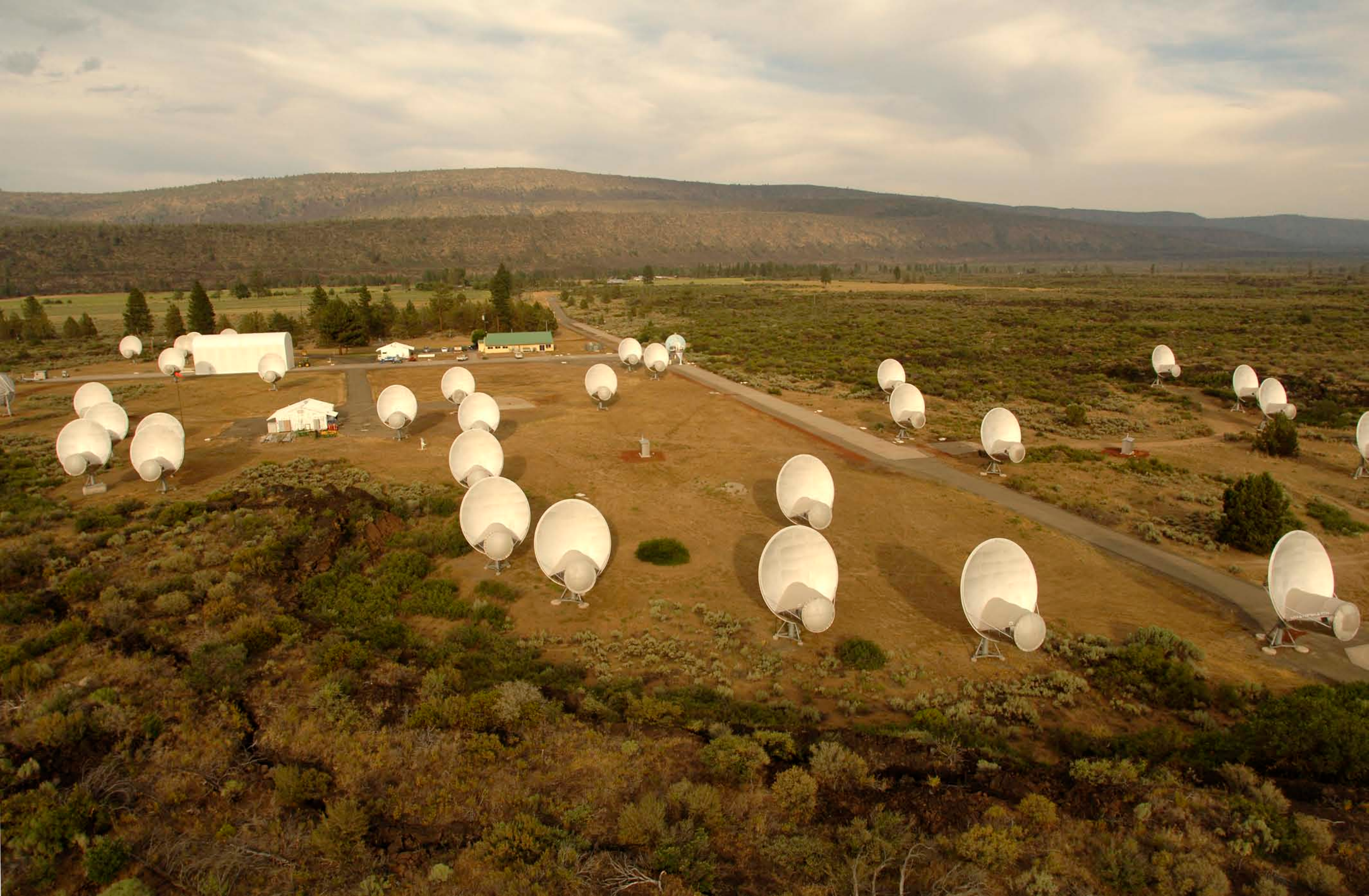}
  \caption{The Allen Telescope Array. Image courtesy of Seth Shostak and the SETI institute.}
 \label{ATA}
}
\end{figure}
\begin{figure}
\center{
 \includegraphics[width=0.4\textwidth]{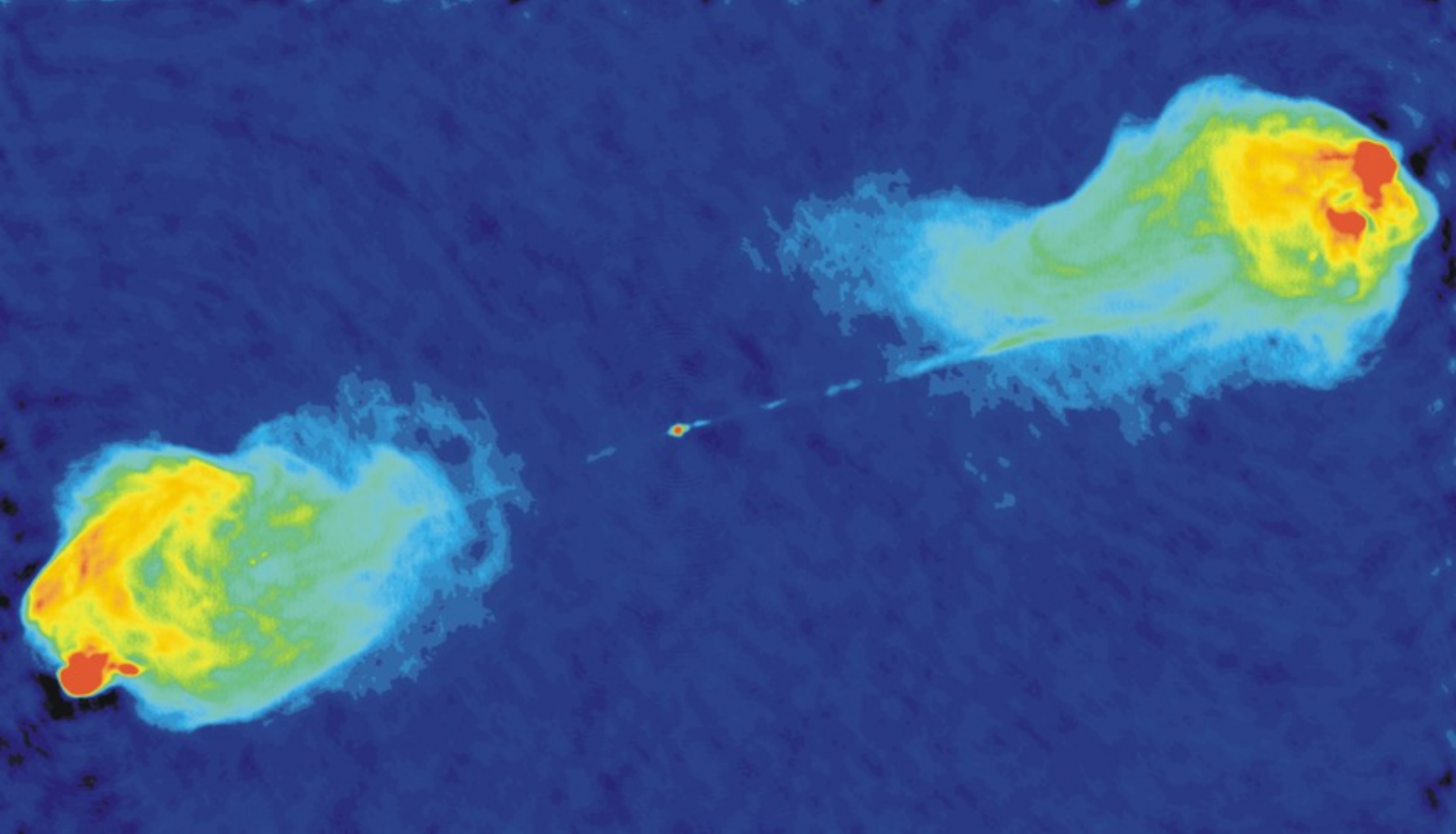}
  \caption{Cygnus A image. False color image of the radio jet and lobes in the
  hyperluminous radio galaxy Cygnus A. Red shows regions with the brightest radio emission, while blue shows regions of
  fainter emission. Image courtesy of NRAO/AUI by Perley et al. \cite{perley84}.  }
 \label{CygA}
}
\end{figure}
In 1962 the principle of synthesis aperture imaging using earth
rotation was proposed by Ryle \cite{ryle62}. Ryle's idea was simple and beautiful. Instead of moving the antennas as
he has been doing for about 15 years, he used the fact that the earth rotates to generate the synthetic aperture. This quickly
became the main operating mode of radio interferometers. However,
imaging using earth rotation synthesis radio telescopes is an ill-posed problem due
to the irregular sub-Nyquist sampling of the Fourier domain.
This sub-sampling results in aliasing inside the image due to the high sidelobes of the
array response. To solve this problem we need to remove the effect of the instrumental response
from the image (a process known as deconvolution) and to compensate for inaccuracies in the array response
(known as self calibration, but it has many similarities to blind deconvolution).
It is important to understand that the improved imaging capability is a result of better equipment in conjunction with
new imaging techniques. Each generation of radio telescopes involved significant hardware development effort.
However, exploiting the hardware capabilities requires a constant improvement in imaging and self calibration to match
the receiver sensitivity. Figure \ref{CygA} (By Perley et al. \cite{perley84})presents the outcome of imaging and
self calibration applied to an image of Cygnus A. It is the first discovery of the radio jets going from the center all
the way to the external radio lobes. Even though Cygnus A has been observed for many years (since Reber's time)
it is the image formation and self calibration algorithms that allowed the discovery of the radio jets.

Over the last 40 years many deconvolution techniques have been developed
to solve this problem. The basic idea behind a
deconvolution algorithm is to exploit a-priori knowledge about the
image. The first algorithm and the most popular of these techniques
is the CLEAN method proposed by H\"{o}gbom \cite{hogbom74}. The
maximum entropy algorithm (MEM) with various entropy functions was
proposed in \cite{frieden72}, \cite{gull78}, \cite{ables74} and
\cite{wernecke77} and the current implementation by Cornwell and
Evans \cite{cornwell85} is the most widely used. Beyond these two
techniques there are several extensions in various directions:
extensions of the CLEAN algorithm to support multi-resolution and
wavelets as well as non co-planar arrays and multiple wavelengths
(see the overview paper \cite{rao2009}). MEM
techniques have been also extended to take into account source
structure through the use of multiresolution and wavelet based
techniques \cite{pantin1996}. Global non-negative least squares was
proposed by Briggs \cite{briggs95}, matrix based parametric imaging
such as the Least Squares Minimum Variance Imaging (LS--MVI) and maximum likelihood based techniques
in \cite{leshem2000a} and \cite{bendavid08}
and sparse $L_1$ reconstruction in \cite{levanda08} and \cite{wiaux09}. Source
modeling is an important issue and various techniques to improve
modeling over simple point source models by using shapelets,
wavelets and Gaussians \cite{reid2006} have been implemented.
A more extensive overview of classical techniques and
implementation issues is given in \cite{thompson86}
or \cite{taylor99}.

Better performance analysis of imaging as well as self calibration
techniques is one of the major challenges for the signal processing
community. This is likely to become a more critical problem for the
future generation of radio interferometers that will be built in the
next two decades such as the square kilometer array\footnote{http://www.skatelescope.org/} (SKA),
the Low Frequency Array\footnote{http://www.lofar.org/p/astronomy.htm}
(LOFAR), the Allen
Telescope Array\footnote{http://ral.berkeley.edu/ata/} (ATA, see
figure \ref{ATA}), the Long Wavelength Array\footnote{http://lwa.unm.edu/} (LWA)
and the Atacama Large Millimeter Array\footnote{http://www.almaobservatory.org/index.php} (ALMA).
These radio-telescopes will be composed of many stations (each
station will be made up of multiple antennas that are combined using
adaptive beamforming). These radio-telescopes will have
significantly increased sensitivity and bandwidth, and some of them
will operate at much lower frequencies than existing radio
telescopes. Improved sensitivity will therefore require a much
better calibration, the capability to perform imaging with much
higher dynamic range in order to reduce the effect of the residuals
of powerful foreground sources inside and outside the field of view and better
handling of non-coplanar arrays.

The structure of the paper is as follows: We begin with a description of the basic imaging equation and a
parametric reformulation of the problem.
The basic approaches to imaging, assuming a calibrated array are
described next. We then describe some modifications of the
parametric imaging techniques that make it possible to combine
imaging and calibration through semi-definite programming. We end up with some simulated and measured examples.
\section{The imaging equations}
This section reviews the basic principles of radio astronomy
following Taylor et al. \cite{taylor99}. In radio astronomy we
observe the radio waves emitted from space. Since the source is far away,
the received electromagnetic
field intensity distribution can be observed only in an angular
direction (no information regarding the intensity distribution in
the radial direction). Defining the \emph{celestial sphere} as the
maximal sphere that contains no radiating sources, the observed
intensity is the projection of the source intensity on the celestial
sphere. For simplicity we will deal with a quasi monochromatic wave
at frequency $\nu$ (the general case can be easily derived by a
linear combination of quasi monochromatic waves). The electric field
at location ${\bf r}$  is given by:
\begin{equation}
\label{eq:electric_field_def}
 E_\nu(\vr) = \int \epsilon_\nu({\bf q}) \frac{ e^{2\pi \jmath \nu
\left| \bf{q} - \vr\right |/c}} {\left| \bf{q} - \vr \right |} {\rm
d} S
\end{equation}
where  $\epsilon_{\gn}(\bf{q})$ is the electric field at location
$\bf{q}$ (on the celestial sphere), ${\rm d} S $ is surface area on
the sphere and the integration is done over the entire sphere and $c$ is the speed of light.
\begin{figure}
\label{fig: correlators and coordinate system}
 \subfigure[] {
\label{subFig:correlator}
  \includegraphics[width=0.23 \textwidth]{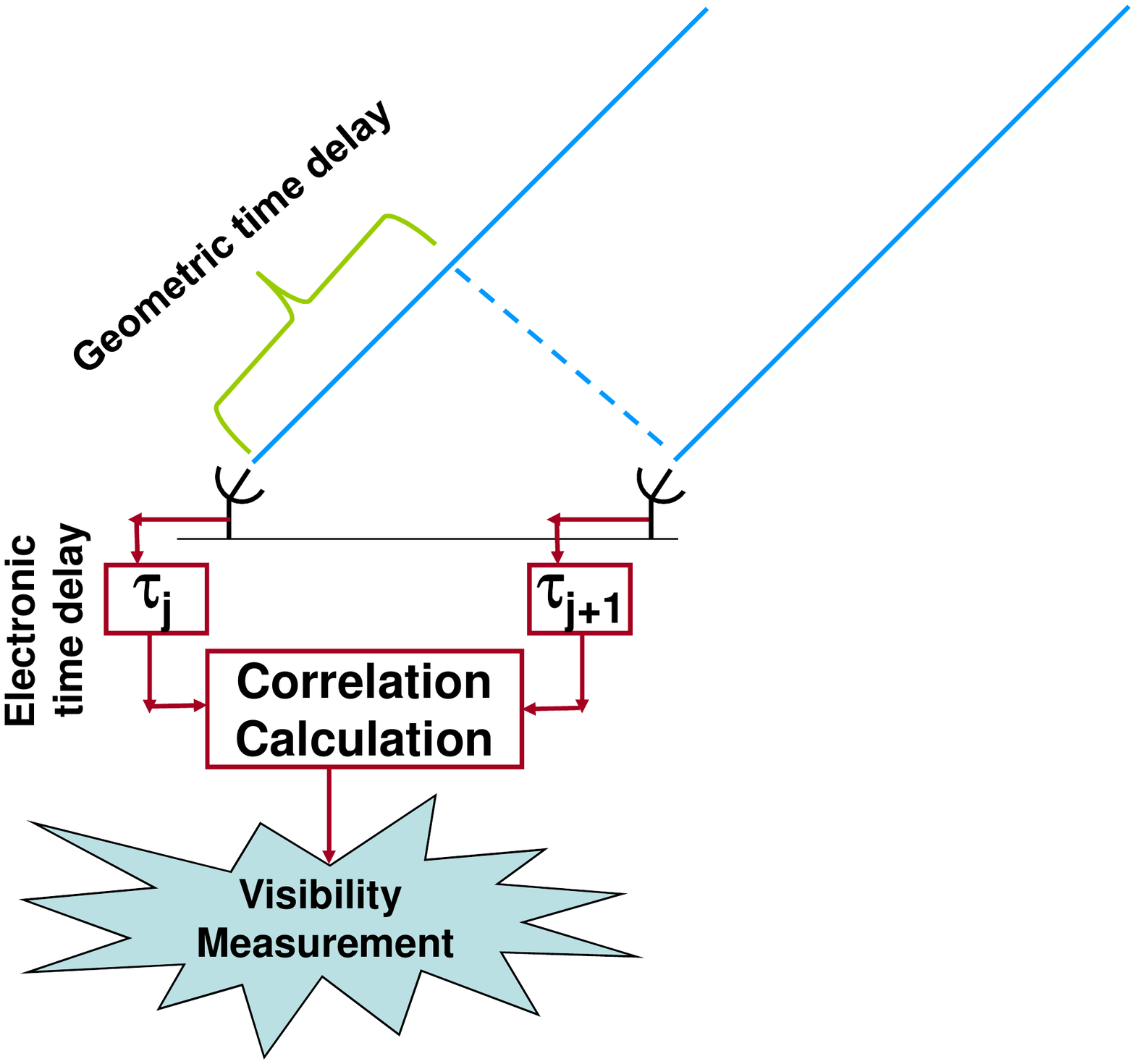}}
    \subfigure[] {
\label{subFig:coordinate_system_def}
 \includegraphics[width=0.23 \textwidth]{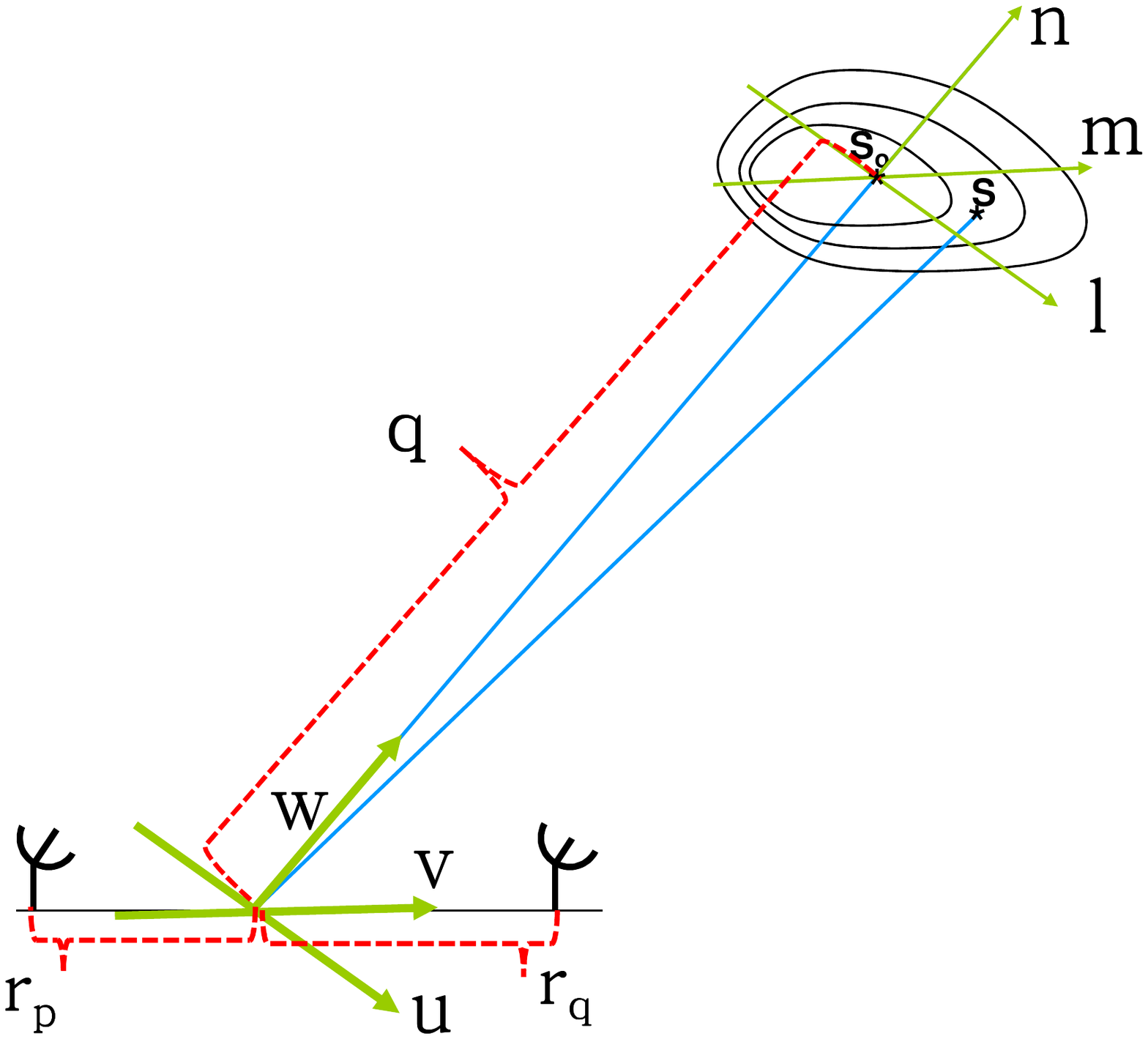}}
 \caption{Measurement setting. \ref{subFig:correlator} - The visibility is the measurement of spatial
 correlation between the electric field of antenna pairs. The geometric delay of the wave that propagates from the source to the two antennas is compensated for by an electronic delay.
 \ref{subFig:coordinate_system_def} - A distant source is observed by an antenna pair.
 The baseline connecting the two antennas is the origin of the  $(u,v,w)$ coordinate system.
 The $w$ axis points from the baseline toward the source reference point. $(u,v)$ are
perpendicular to $w$ and selected according to the Earth's
orientation. $(l,m,n)$ is a unit vector in the $(u,v,w)$ system
pointing toward a specific location in the source (at the source
reference point $S_o$, $l = 0,m=0$ ) and $n = \sqrt {\left( l^2+m^2
\right)}$.}
\end{figure}
%
%
%
For two antennas observing a distant source (receiving the electric
field emitted by the source) there is a geometrical delay in one of
the antennas relative to the other antenna derived from the source
observation angle (see Figure \ref{subFig:correlator}); if the
geometric delay is compensated by an electronic delay, the electric
field received in one antenna should be highly correlated with the
electric field received by the other antenna. The spatial coherency
of the electric field for two antennas located at ${\bf r}_1$ and
${\bf r}_2$ is given by
\begin{equation}
\label{eq:coherency_def} V_\nu(\bf {r_1},\bf {r_2})= \langle
\bf{E}_\nu({\bf r}_1) \bf{E}^*_\nu({\bf r}_2)\rangle.
\end{equation}
where $\langle \rangle$ stands for the expectation value.
Substituting (\ref{eq:electric_field_def}) into
(\ref{eq:coherency_def}) and taking into account the large distance
of the source; i.e. $\left| \bf {q-r} \right| \approx \left| \bf {q}
\right|$ and that the electric field is spatially incoherent (i.e.,
$\langle\epsilon_\nu ({\bf q}_1) \epsilon^*_\nu ({\bf q}_2) \rangle
= 0 \phantom a \forall {{\bf q}_1} \neq {{\bf q}_2}$) we get
\begin{equation}
\label{eq:visibility_def_int} V_\nu({\bf r}_1,{\bf r}_2) = \int
I_\nu(s) e^{-2 \pi \jmath \nu {\bf s} ({\bf r}_1 - {\bf r}_2)/c} d\Omega,
\end{equation}
where $I_\nu({\bf s}) \equiv \langle \epsilon_{\gn}(\vs) ^2 \rangle$ is the
source intensity at direction $\bf s$ on the sphere ($\bf s \equiv
\frac{\bf q}{|\bf q|}$), and $d \Omega = \frac{dS} {|{\bf q}|^2}.$
%
 Representing (\ref{eq:visibility_def_int}) in the $(u,v,w)$
coordinate system, for many astronomical observations (e.g., planar
arrays, or small field of view imaging) we obtain
\begin{equation}
\label{eq:visibility_uv} V_\nu(u,v) = \int \int I_\nu (l,m) e^{-2
\pi \jmath (ul+vm)} {\rm d} l {\rm d} m .
\end{equation}
The visibility is the Fourier transform of the source intensity;
therefore the inverse relation holds :
\begin{equation}
\label{eq:intensity_lm_eq_vis} I_\nu(l,m) = \int \int V_\nu (u,v)
e^{2 \pi \jmath (ul + vm)} {\rm d} u {\rm d} v.
\end{equation}

When the co-planar approximation does not hold, equation (\ref{eq:visibility_uv})
takes the more complicated form
\begin{equation}
\label{eq:visibility_uvw_non_coplanar}
 V_\nu(u,v,w)  = \int \int
\frac{1}{n} I_\nu (l,m)
 e^{-2 \pi \jmath \left[ul+vm+w(n-1)\right]} {\rm d} l {\rm d} m,
\end{equation}
where
\begin{equation}
\label{eq:visibility_normalization_factor}
 n \equiv \sqrt{1-\ell^2-m^2}.
\end{equation}

For a source with visibility measurements covering the entire
$(u,v)$ domain, the source image is perfectly computed by the
Fourier inversion of the visibility. In practice, only a small part
of the $(u,v)$ domain is measured by sampling the existing antenna
pair baselines as they change with the Earth's rotation relative to
the $(u,v)$ coordinates (at time $t_k$ two antennas $p$ and $q$
measure a single visibility point in the $(u,v)$ domain at
$(u_{pq}^k,v_{pq}^k)$) . This set of samples is known as
the $(u,v)$ coverage of the radio telescope. This coverage is
determined by many factors such as the configuration in which the
individual receptors (telescopes or dipole) are placed on the
ground, the minimal and maximal distance between antenna pairs, the
time difference between consecutive measurements and the total
measurement time and bandwidth. An example of the $(u,v)$ coverage
for a simulated radio telescope (East west array with 14 antennas
logarithmically spaced from $\lambda$ to $200 \lambda$, observation
time of 12 hours) is shown in Figure \ref{fig:uvw_coverage}. The
sampled points in the $(u,v)$ plane are a collection of ellipses.
The sampling effect on the resulting image is shown in Figure
\ref{subFig:perfect_grid_img} and
\ref{subFig:realistic_uv_coverage_perfect_gird}. Figure
\ref{subFig:perfect_grid_img} depicts an image of visibility data
measured over a dense and uniform grid in the $(u,v)$ plane (all grid points in the $(u,v)$
plane were sampled). Figure
\ref{subFig:realistic_uv_coverage_perfect_gird} presents the same
data with a more realistic $(u,v)$ sampling. The image with the
partial (and more realistic) measurement set is blurred, distorted
and noisy. Let $S(u,v)$ be the sampling function ($S(u,v) = 1$ for
each measured $(u,v)$ pair and $S(u,v) = 0$ otherwise). We obtain
that the inverse direct Fourier transform of the measured
visibility, known as the \emph{dirty image $I_D$} is given by:
\begin{figure}
\center{
\includegraphics[width=0.4\textwidth]{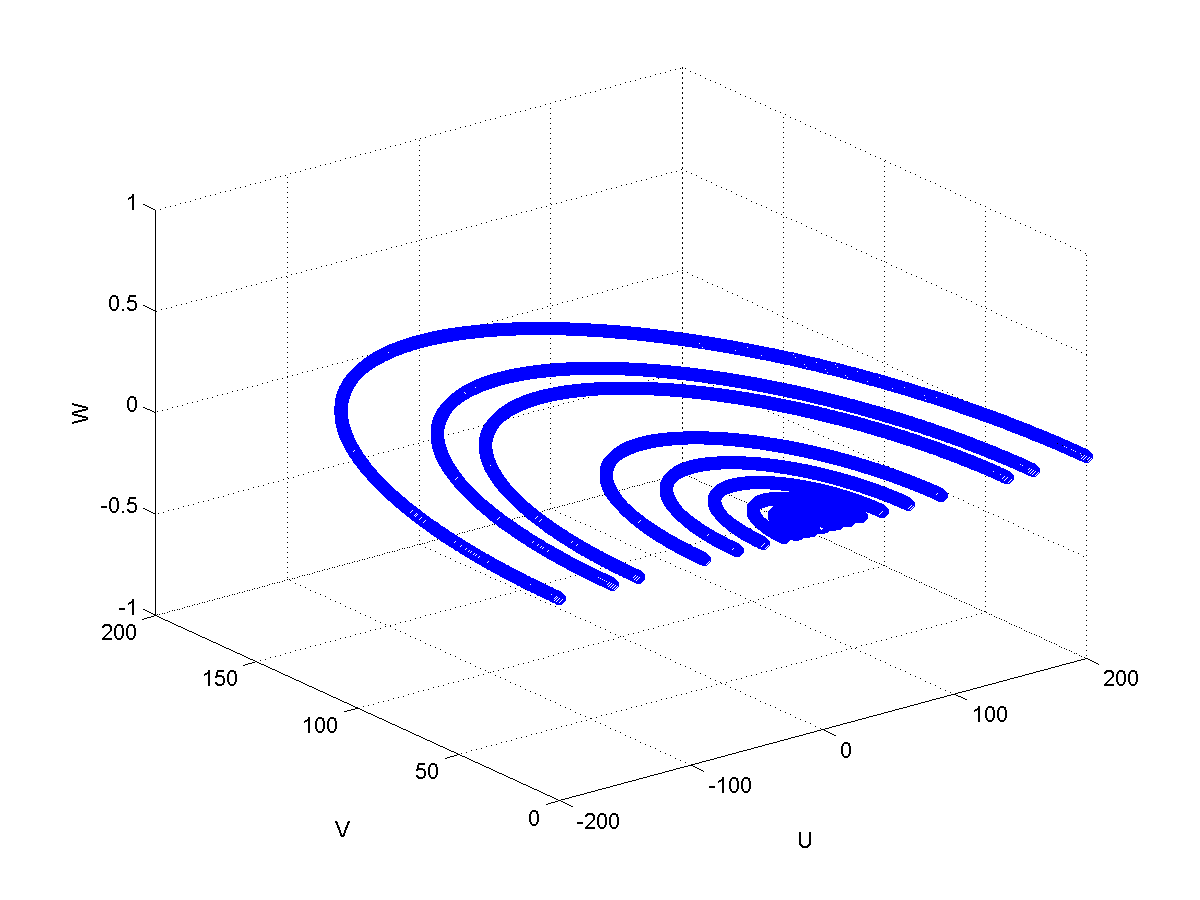}
  \caption{$(u,v)$ coverage of a simulated East-West radio telescope with 14 antennas logarithmically
  spaced
  with baselines up to 200$\lambda$. Observations are made every 6 minuets for a duration of 12 hours.
  From each antenna pair we get an ellipse in the $(u,v)$ domain. $u$ and $v$ are in $\lambda$ units.}
\label{fig:uvw_coverage}
}
\end{figure}

\begin{figure}
\label{fig: grid_exp}
\center{
 \subfigure[] {
\label{subFig:perfect_grid_img}
  \includegraphics[width=0.3\textwidth]{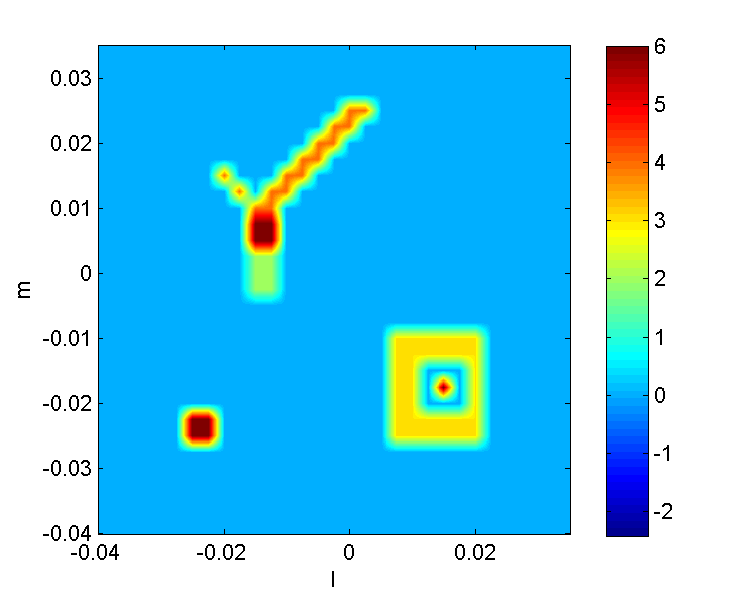}}
    \subfigure[] {
\label{subFig:noisy_grid_img}
 \includegraphics[width=0.3\textwidth]{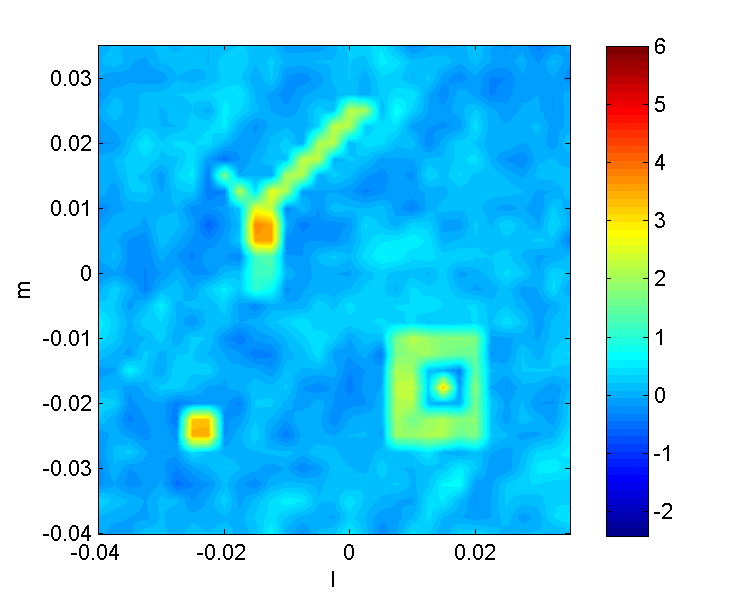}}
 \subfigure[]{
 \label{subFig:realistic_uv_coverage_perfect_gird}
 \includegraphics[width=0.3\textwidth]{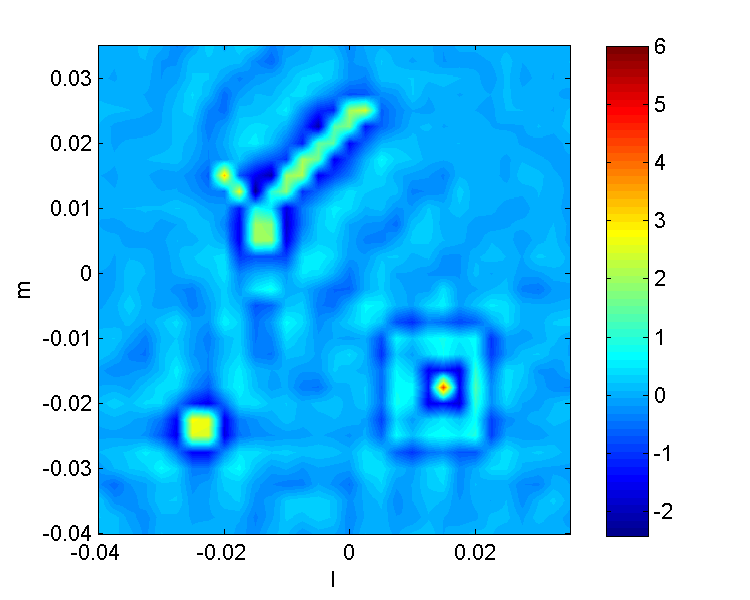}}
   \caption{Illustration of sampling and gridding effects.  \ref{subFig:perfect_grid_img}: Fourier transform
   of  visibility data on a perfect grid (visibility measurements location are on the center of the grid cells) with
   complete $(u,v)$ coverage.
   \ref{subFig:noisy_grid_img}: Fourier transform of visibility data with off-grid points
   (the visibility measurements location were chosen randomly) and complete $(u,v)$ coverage,
    demonstrates gridding effect. Features in the image are blurred and distorted.
  \ref{subFig:realistic_uv_coverage_perfect_gird}: Fourier transform of visibility data with perfect grid and incomplete
 $(u,v)$ coverage (of the radio telescope described in
  Figure \ref{fig:uvw_coverage}), demonstrates the sampling effect. }
}
\end{figure}

\begin{equation}
\label{eq:dirty_img_def} I_{D,\nu}(l,m) = \int \int V_\nu (u,v)S(u,v)
e^{2 \pi \jmath (ul + vm)} {\rm d} u {\rm d} v.
\end{equation}
The instrument point spread function which is also known as the dirty beam is defined
by:
\begin{equation}
\label{eq:dirty_beam_def} B(l,m) \equiv \int \int S(u,v) e^{2 \pi \jmath
(ul + vm)} {\rm d} u {\rm d} v.
\end{equation}
By the convolution theorem, the dirty image is the convolution of the true source
intensity (\ref{eq:intensity_lm_eq_vis}) and the dirty beam
(\ref{eq:dirty_beam_def}):
\begin{equation}
\label{eq:dirty_img_conv} I_{D,\nu} = I_\nu * B
\end{equation}
This is the reason why image reconstruction algorithms in radio
astronomy are often referred to as \emph{deconvolution} algorithms,
since direct synthesis produces $I_{D,\nu}$, but we  want to obtain
$I_\nu$ by deconvolution with respect to $B$. The dirty image can be
calculated from the measured visibility data according to equation
(\ref{eq:dirty_img_def}), or by using a Fast Fourier Transform (FFT)
to reduce the calculation time and memory resources. In order to use
the FFT, the visibility data must lie on a rectangular equally
spaced grid. This procedure of re-sampling the measured visibilities
on a regular grid is called \emph{gridding}. The weighting is done
by convolving the visibilities with a smooth kernel (this procedure
is also called convolutional gridding). Choice of the gridding
kernel is important and follows from standard interpolation theory.
An illustration of the gridding effect for a rectangular kernel is
shown in Figures \ref{subFig:perfect_grid_img} and
\ref{subFig:noisy_grid_img}. Both images were generated using
simulated visibility data with complete $(u,v)$ coverage. In Figure
\ref{subFig:perfect_grid_img} the visibility data were taken on a
perfect grid (all visibility measurements were located on the center
of a grid cell). In Figure \ref{subFig:noisy_grid_img} the location
of the visibility measurements was chosen randomly within the cells
in the $(u,v)$ plane). This results in a blurred and distorted
image. For more details on gridding and tapering the reader is
referred to \cite{thompson86} and \cite{taylor99}.

\section{The parametric matrix formulation of the image formation problem }
We now describe an alternative formulation of the image formation
problem. In this formulation imaging is viewed as a parameter
estimation problem, where the locations and powers (and possibly
polarization parameters and frequency dependence of power) are the
unknown parameters. This formalism was first proposed in
\cite{leshem2000b}  and \cite{leshem2000a} to allow for the
introduction of interference mitigation techniques in the imaging
process. It was extended to non co-planar array and polarimetric
imaging in \cite{bendavid08}. This formulation also allows easy inclusion of space dependent
calibration parameters \cite{vandertol2009}. Assume that the observed image is a
collection of $D$ point sources, i.e.,
\begin{equation}
\label{eq:intensity_discerete_def} I_\nu(l,m) = \sum_{d =
1}^{D} I_\nu(l,m) \delta(l - l_{d}) \delta(m - m_{d}).
\end{equation}
Since $(u,v)$ are the baseline coordinates (i.e. $u \equiv u_{ij}^k
= u_i^k - u_j^k$ and $v \equiv v_{ij}^k= v_i^k - v_j^k$), the
visibility (\ref{eq:visibility_uv}) can be rewritten as
\begin{equation} \label{visibility_uvt_descrete}
V_\nu(u_{ij}^k,v_{ij}^k) = \sum_{d = 1}^{D} I_\nu (l_d,m_d)
e^{-2 \pi \jmath(u_{ij}^k l_d+v_{ij}^k m_d)}.
\end{equation}
where $k$ denotes the measurement time $t_k$. Selecting a (time
varying) reference point at one of the antennas $(u_0^k,v_0^k)$ and
manipulating (\ref{visibility_uvt_descrete}) yields \beq \bea{lcl}
\label{eq:visibility_with_ref_point} V_\nu(u_{ij}^k,v_{ij}^k) & = &
\sum_{d= 1}^{D} e^{-2 \pi \jmath(u_{i,0}^k  l_d+
v_{i,0}^k m_d)} I_\nu (l_d,m_d) \\
& & e^{2 \pi \jmath(u_{j,0}^k l_d+ v_{j,0}^k m_d)} \ena \eeq

\begin{figure}
\includegraphics[width=0.4\textwidth]{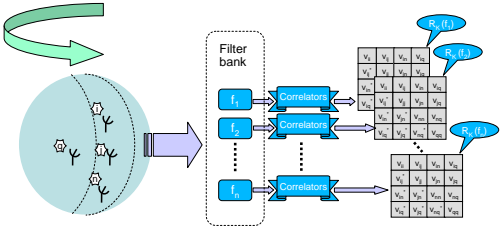}
\caption{The correlation matrices for an observation in the multi
frequency case. The correlation matrices are calculated
for each frequency separately (on filtered measurements from the
antennas).}
 \label{fig: filter bank}
\end{figure}

We define the $k$'th measurement correlation matrix $\mR_k$ by:
\begin{equation}
 (\mR_k)_{ij}  \equiv
 V_\nu(u_{ij}^k,v_{ij}^k)
\end{equation}
The correlation matrix is illustrated in Figure \ref{fig:
corr_mat_illustration}, for a single frequency bin. Cell
$R_{ij}^k$ of the correlation matrix is the visibility measurement
at time $t_k$ from antenna pair $(i,j)$. The size of the correlation matrix $\mR_k$ is
$p\times p $ where $p$ is the number of antennas in the array.
The autocorrelation of each antenna is also used (the diagonal of the correlation
 matrix). When an observation uses more than a single frequency
 bin, each correlation matrix is computed using a single bin as illustrated in Figure \ref{fig: filter bank}.

\begin{figure}
\includegraphics[width=0.4\textwidth]{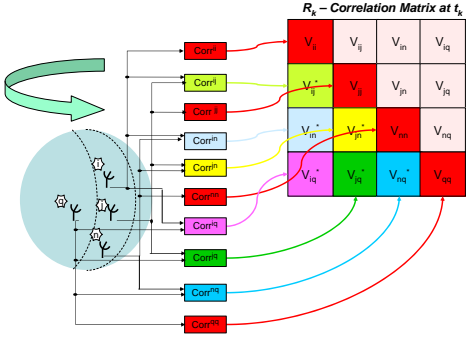}
\caption{The correlation matrix (for the simple case of a single
frequency) in a specific time measurement $t_k$ is built from the
visibility measurement of antenna pairs} \label{fig:
corr_mat_illustration}
\end{figure}

Let the Fourier component vector at time $t_k$ be:
\begin{equation}
\label{eq:ak vec def}
 {\bf a}_k(l,m) \equiv  \left(
  \begin{array}{c}
    e^{-2\pi \jmath (u_{1,0}^k l + v_{1,0}^k m)} \\
    \vdots \\
    e^{-2\pi \jmath (u_{p,0}^k l + v_{p,0}^k m)} \\
  \end{array}
\right) ,
\end{equation}
and let the Fourier component matrix at time $t_k$ be
\begin{equation}
\label{eq:Ak matrix def}
 \mA_k  \equiv \left[
  \begin{array}{ccc}
    {\bf a}_k(l_1,m_1) ,& \ldots & ,{\bf a}_k(l_d,m_d) \\
  \end{array}
\right].
\end{equation}
Define the point source intensity matrix by :
\begin{equation}
   \label{eq:B point source matrix def}
\mB \equiv  \left[
           \begin{array}{ccc}
             I(l_1,m_1) &  & 0 \\
              & \ddots &  \\
             0 &   & I(l_{D},m_{D}) \\
           \end{array}
         \right].
\end{equation}

Using   (\ref{eq:ak vec def})-(\ref{eq:B point source matrix def}) equation (\ref{eq:visibility_with_ref_point}) can be rewritten as

\begin{equation}
\label{eq:corr_mat_vis__func_of_intens_in_param_form} \mR_k = \mA_k \mB
\mA_k^H.
\end{equation}
Matrix equation
(\ref{eq:corr_mat_vis__func_of_intens_in_param_form}) is the
parametric form of the classical equation (\ref{eq:visibility_uv}).
It will allow us to consider the problem as an estimation problem,
where we observe a set of measured covariance matrices which depend
smoothly on the unknown source and instrument calibration
parameters, as well as receiver noise. Using this formulation we can
easily use well-known techniques from estimation theory (such as
maximum a-posteriori, ML, MVDR and robust techniques) to solve the
image formation problem. It also enables a simple extension to the
non-coplanar array case as well as polarimetric imaging and
multi-frequency synthesis, where sources have frequency dependent
(parametrically known) characteristics. The classical dirty image
(\ref{eq:dirty_img_def}) can be rewritten as
\begin{equation}
\label{eq:dirty_img_func_of_corr_in_param_form} I_D(l,m) =
\frac{1}{K} \sum_k \va_k^H(l,m) \mR_k \va_k(l,m).
\end{equation}
Note that this is identical to the mean power output of a classical
beamformer oriented towards direction $(l,m)$. More
realistically, the antenna response varies slightly between
different antennas and there is an additional noise per antenna. The
antenna response can be measured prior to the observation and taken
into account in the model. Since the noise in two antennas is
independent, the noise correlation matrix is diagonal. Denoting by
$\gamma_{i,k}$ the unknown complex gain of antenna $i$ at
observation time $t_k$ and by $\sigma^2$ the noise variance, the
correlation matrix now becomes:
 \beq
  \label{eq:corr mat with gamma}
  \mR_k= \mgG_k \mA_k \mB \mA_k^H \mgG_k^H  + \sigma^2 \mI
  \eeq
  where
  \[
  \mgG_k \equiv \left[
                      \begin{array}{ccc}
                        \gamma_{1,k} &   & 0 \\
                         & \ddots &  \\
                        0 &  & \gamma_{p,k} \\
                      \end{array}
                    \right]
\]
Estimation of the $\gamma_{i,k}$ is discussed in a companion paper
by Wijnholds et al. \cite{wijnholds2009}. Note that typically
$\gamma_{i,k}$ varies slowly so it can be assumed to be constant
over multiple times. Similarly, the non-coplanar array case is given
by replacing $\va_k$ and $\mB$ by
\beq
\label{eq: param non coplanar_array}
\bea{ll}
  \va_k(l,m) &\equiv \\
    \left[e^{-2\pi \jmath (u_{1,0}^k l + v_{1,0}^k m + w_{1,0}^k n)},...,
   e^{-2\pi \jmath (u_{p,0}^k l + v_{p,0}^k m + w_{1,0}^k n)}\right]^T
\ena
\eeq
and
\beq
\mB
 \equiv  \left[
           \begin{array}{ccc}
             \frac{I(l_1,m_1)}{\sqrt{1-\ell_1^2-m_1^2}} &  & 0 \\
              & \ddots &  \\
             0 &   & \frac{I(l_{D},m_{D})}{\sqrt{1-\ell_{D}^2-m_{D}^2}} \\
           \end{array}
         \right].
\eeq The radio imaging problem can now be reformulated as follows:
Given a set of measured covariance matrices $\mRh_1,...,\mRh_K$
estimate the parameters $\vs_1,...,\vs_{D}$, $I(\vs_1),...,I(\vs_D)$ and the
calibration matrices $\mgG_k:k=1,...,K$. Note that (\ref{eq:corr mat
with gamma}) can be easily generalized to deal with direction
dependent calibration parameters, polarized sources as well as
multi-frequency synthesis. All that we need to change is the source
and the calibration parametric model by simple adaptation of
(\ref{eq:corr mat with gamma}). The parametric approaches described
in this paper can be applied uniformly to all these problems.
However, for simplicity we will focus on the calibrated array case.
\section{Classical and parametric approaches based on sequential source
removal} Many algorithms in radio astronomy are based on sequential
source removal. The most commonly used  is the CLEAN algorithm
originally proposed by H\"{o}gbom \cite{hogbom74}. These iterative
algorithms assume that the observed field is a collection of sources
with simple structure. CLEAN assumes that the sources are point
sources. During each iteration a single point source is estimated
and removed from the data. The reconstructed image is the collection
of all point sources with their estimated power convolved with an
ideal reconstruction beam (usually an elliptical Gaussian fitted to
the central lobe of the dirty beam) . The general structure common
to all the sequential source removal algorithms is described in
Table \ref{tab:gen source remove algo flow}. The algorithms differ
from each other by the exact definition of the dirty image used, the
way the point source is removed from the image (either in the image
domain after gridding or in the visibility domain), the intensity
estimation method of the point sources and the exact modeling of the
sources (point source, Gaussian, wavelet coefficients, shapelets,
etc.). Some versions like the Cotton-Schwab technique estimate
multiple sources based on the same dirty image. This significantly
accelerates the algorithm, since the number of Fourier transforms of
the image is reduced.

We describe two sequential source removal algorithms. The first is
the CLEAN algorithm and the second is a parametric estimation based
algorithm known as LS-MVI.
\begin{table}
  \begin{tabular}{|p{0.45\textwidth}|}
  \hline
  \textbf{Initialization}: \\
  \quad $\bullet$ Calculate the dirty image $I_D$ according to measured
  visibilities. \\
  \quad $\bullet$ Calculate the reconstruction beam $B_{rec}$ for later use. \\
  \textbf{While stopping criteria not met}:\\
  \quad $\bullet$ Find the brightest location in the dirty image $(l_i,m_i)$. \\
   \quad \quad This is the location of a new point
  source. \\
  \quad $\bullet$ Estimate the new point source intensity $\gl_i$. \\
  \quad $\bullet$ Add the new point source to the source list. \\
   \quad  \quad (with the estimated intensity). \\
   \quad $\bullet$ Remove the new source response from the data \\
    \quad  \quad (both the dirty image and the visibility
   measurements).\\
 \textbf{Finalize}:\\
  \quad $\bullet$ Calculate the reconstructed image $I_{rec}$ \\
   \quad  \quad by convolving
  the source list with the reconstruction beam. \\

   \hline

\end{tabular}
\caption{Generic source removal algorithm flow}
 \label{tab:gen source remove algo flow}
\end{table}
\subsection{The CLEAN Algorithm}
The CLEAN algorithm assumes that the observed field of view is
composed of point sources. Since the image of a point source is
given by the convolution of the point source and the dirty beam
(\ref{eq:dirty_img_conv}), CLEAN iteratively removes the brightest
point source from the image until the residual image is noise-like.
There are several variants of CLEAN
(\cite{hogbom74},\cite{cornwell2008a},\cite{Clark1980},\cite{Schwab1984}).
The CLEAN algorithm is implemented either in the image or in the
visibility domain. In each iteration the brightest point in the
dirty image (equation (\ref{eq:dirty_img_def}) ) is found
(position and strength) and added to a point source list. A fraction
of it ($\gamma$, $0<\gamma<1$) is removed from the dirty image. The
$\gamma$ parameter is called the loop gain and is usually taken to
be 0.1-0.2. The iterations continue until the residual image is
noise like. The subtraction can be done either in the image domain
or in the visibility domain. The visibility domain CLEAN is
more accurate since we are not limited to pixel resolution. The algorithm flow for ungridded visibility domain
CLEAN is summarized in Table \ref{tab:basic clean algo flow}.

\begin{table}
  \begin{tabular}{|p{0.45\textwidth}|}
  \hline
  \textbf{Initialization}: \\
  \quad $\bullet$ Calculate $I_D$ (eq. \ref{eq:dirty_img_def}).\\
  \quad $\bullet$ $i = 0$. \\
   \quad $\bullet$ $B_{\rm rec} = {\rm Gaussian}$. \\
  \textbf{While $I_D$ is not noise-like}:\\
  \quad $\bullet$ $(l_i,m_i) = {\rm arg} \max I_D(l,m)$. \\
  \quad $\bullet$ $\lambda_i = I_D(l_i,m_i)$.\\
  \quad $\bullet$ For all $p,q,k$: \\
  \quad \ \ \ \ $\tiny {V(u_{p,q}^k,v_{p,q}^k) = V(u_{p,q}^k,v_{p,q}^k) - \gamma \lambda_i e^{-2 \pi j \left[u_{p,q}^k l_i + v_{p,q}^k m_i   \right]}}$.\\
  \quad $\bullet$  Update $I_D$ (eq. \ref{eq:dirty_img_def}).\\
   \quad $\bullet$ $ i = i + 1$. \\
 \textbf{Finalize}:\\
  \quad $\bullet$ $I_{\rm rec} = I_D + \sum_i \gamma \lambda_i B_{\rm rec}(l - l_i, m -
  m_i)$. \\
   \hline
\end{tabular}
  \caption{Visibility domain CLEAN algorithm flow}
  \label{tab:basic clean algo flow}
\end{table}

\begin{figure}
\begin{tabular}{|cc|}
  \hline
   & \\
   \includegraphics[width=0.2\textwidth]{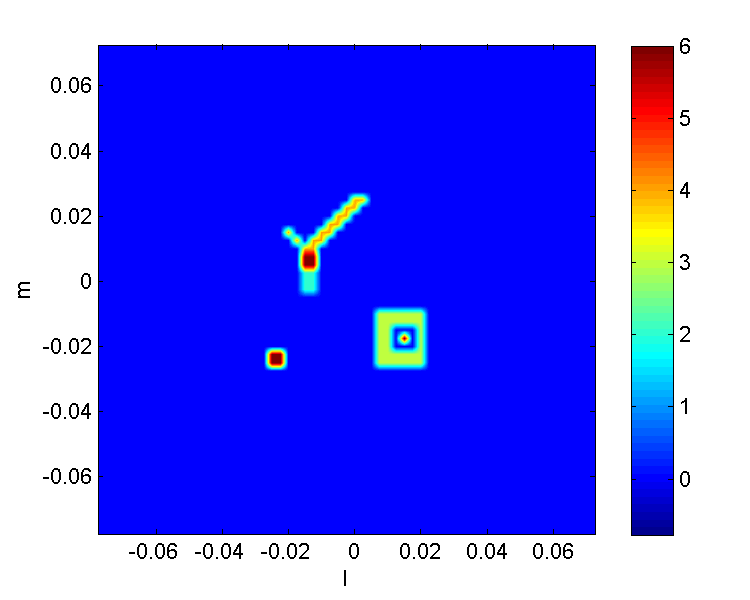} &   \includegraphics[width=0.2\textwidth]{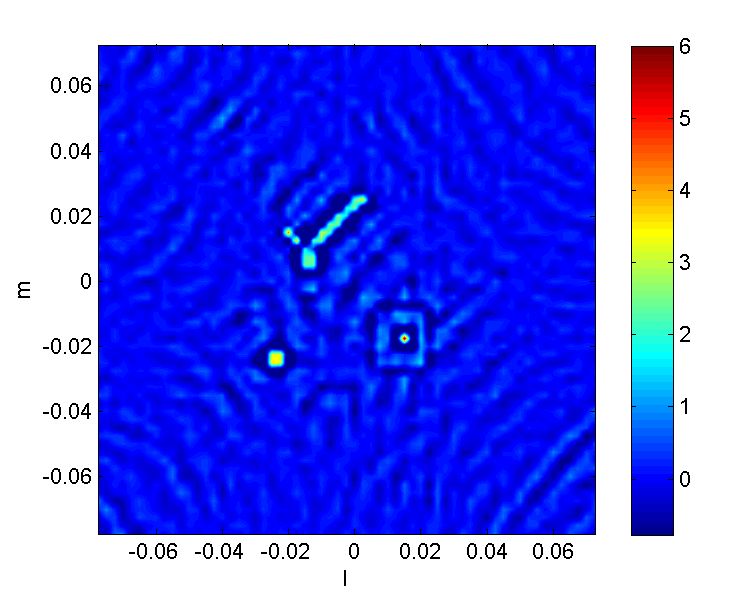}\\
   {\tiny  (a) Original Image} & {\tiny (b) Initial Dirty Image} \\
    \includegraphics[width=0.2\textwidth]{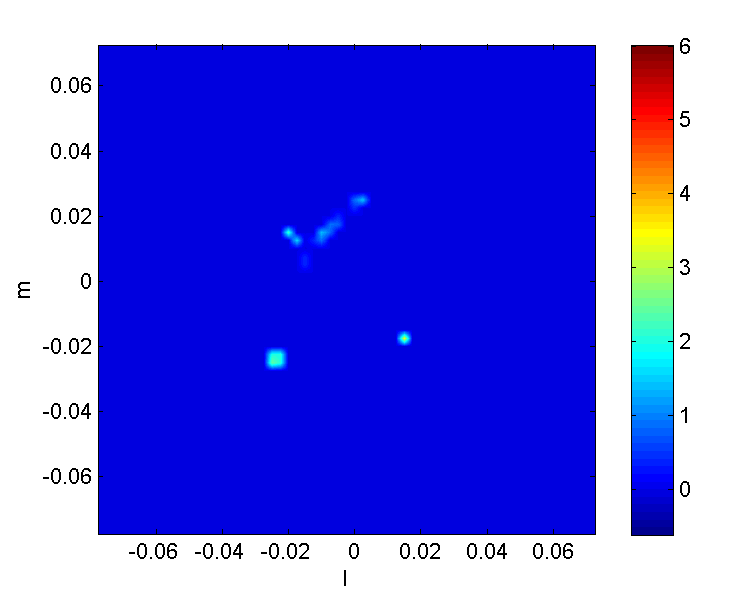} &   \includegraphics[width=0.2\textwidth]{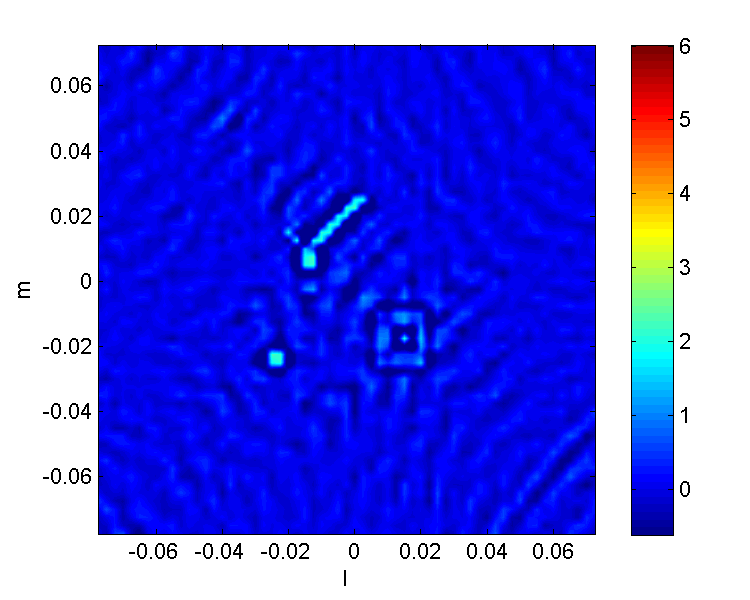}\\
    {\tiny (c) Reconstructed Image - 50 iter }& {\tiny (d) Residual Image - 50 iter} \\
     \includegraphics[width=0.2\textwidth]{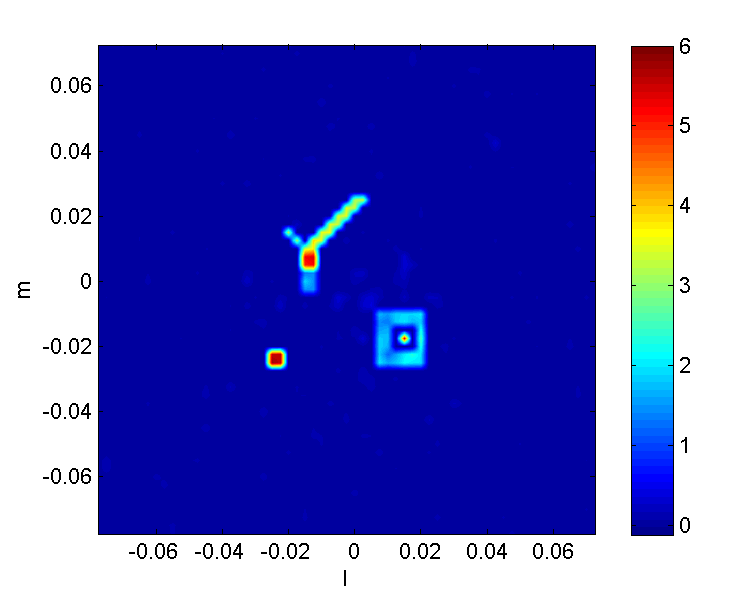} &   \includegraphics[width=0.2\textwidth]{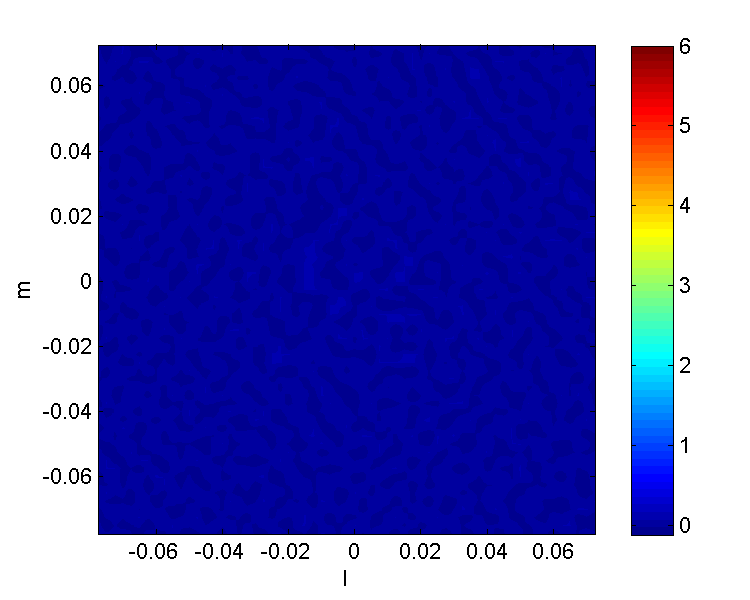}\\
    {\tiny (e) Reconstructed Image - 25000 iter }& {\tiny (f) Residual Image - 25000 iter} \\
   &  \\
  \hline
\end{tabular}
\caption{CLEAN steps for a simulated image. (a) the original image.
(b) the initial dirty image.
(c) and (e) the reconstructed image after 50 and 25000
CLEAN iterations respectively. (d),(f) the residual dirty
images after 50 and 25000 iterations respectively. In general,
the sources are nicely reconstructed except from the square ring
extended source. All images share the same color map. All images
were up-sampled by 4 using Matlab basic interpolation.}
\label{fig:clean sim example}
\end{figure}

An illustration of the CLEAN algorithm on a simulated image is
shown in Figure (\ref{fig:clean sim example}). The simulated radio
telescope is the same as in Figure (\ref{fig:uvw_coverage}). The
loop gain used is $0.2$. In every iteration the strongest point
source is found, added to the reconstructed image and subtracted
from the dirty image. The loop gain serves three purposes. First, it
prevents (or at least reduces) the effects of over-estimation of the
power due to sidelobes from other sources. Second, it allows for
interpolation of sources that are located off the grid. Third, it
improves performance with extended sources. However, this limits the
dynamic range of the image. The effect of pixelization and choice of
grid on the dynamic range of the imaging process is further
discussed in \cite{cotton2008}, \cite{voronkov2004}. Improved
versions of CLEAN allow for estimation of location off the grid by
using interpolation, and subtraction of the effect from the
visibility rather than the dirty image. This has the positive effect
of eliminating gridding accuracy effects. Acceleration of the CLEAN
algorithm can be achieved by estimating multiple point sources based
on a single dirty image (major cycle), as well as defining windows
for the search procedure. Practically, defining windows reduces the
size of the search space.

\subsubsection{Clark CLEAN Algorithm}
One of the important variants of CLEAN was proposed by Clark
 in 1980 \cite{Clark1980}. Clark's algorithm main advantage is reduction
of computational load. The algorithm is performed in two cycles, a
major cycle and a minor cycle. A major cycle is constructed by
selecting intensity limit value (according to a histogram of the
dirty image values) and approximating a dirty beam (central patch of the true dirty beam)
to be used during the subsequent minor
cycles. A minor cycle consists of finding the brightest pixel in the
image (i.e. a new point source) and removing a fraction of the point
source response from the dirty image. In principle, the minor cycle
is the same as described in the 'While' loop in Table \ref{tab:basic
clean algo flow}, when the dirty beam used is only the centeral patch of the
full dirty beam (hence computational complexity is significantly
reduced). The inaccuracies caused by working with an approximated
dirty beam are corrected during the major cycle. The Clark algorithm
is performed in the visibility domain instead of the image domain,
yielding a multiplication instead of a convolution for calculating
the point source response.

\subsubsection{Cotton Schwab Algorithm}
Cotton \& Schwab \cite{Schwab1984} developed a variant of the Clark
CLEAN. Like the Clark CLEAN, in the Cotton Schwab CLEAN the
procedure involves major and minor cycles. The main improvements
over the Clark algorithm are that the Cotton Schwab algorithm
calculation is done over the \emph{ungridded} visibility data, thus
avoiding gridding errors, and multi source removal is done
independently in each minor cycle (from different fields). The CLEAN
components from all fields are removed in the major cycle. Working
with the \emph{ungridded} visibility measurement is done using a
measurement list as described in Table \ref{tab:basic clean algo
flow}. An element $V(u_{p,q}^k,v_{p,q}^k)$ of the measurement list
is the measured visibility by an antenna  pair $(p,q)$,
corresponding to a baseline $(u_{p,q},v_{p,q})$ measured at time
$k$.
\subsection{The W-Projection algorithm}
One of the main limitation of the previous technique is the case of non-coplanar arrays
and large field of view. To overcome problems related to non-coplanar arrays the W-projection algorithm has been proposed by Cornwell et al. \cite{Cornwell2008}.

The W-projection algorithm deals with non-coplanar arrays i.e. when
the planar approximation is violated and the imaging equation
is given by equation (\ref{eq:visibility_uvw_non_coplanar}).
Originally Frater and Docherty \cite{Frater1980} showed that a
projection of visibility measurements from a constant $w$ plane to $w=0$ plane can be done.
This corresponds to a radio telescope with antennas arranged
in a plane with a single antenna outside the plane. In this case the
measured visibilities are projected onto $w=0$ plane (real and
imaginary part separately), a deconvolution is performed (such as
CLEAN) and the resulting cleaned images are combined taking the
constant $w$ value into account.

In the general case (projection of any $w$ values) the relation
between $V(u,v,w)$ and $V(u,v,w=0)$ is given by
\begin{equation}
\label{eq:w projection visibility projection}
 V(u,v,w) =
\tilde{G}(u,v,w)*V(u,v,w=0)
\end{equation}
where
\begin{eqnarray}
  G(l,m,w) &\equiv& e^{-2 \pi j \left[
w\left(\sqrt{1-l^2-m^2} - 1
\right) \right]} \\
\nonumber & \approx& e^{\pi j \left[w \left( l^2 + m^2 \right)
\right]} \\
\nonumber \tilde{G}(u,v,w) &=& \frac{j}{w} e^{-\pi j \left[
\frac{u^2+v^2}{w}\right]} ,\label{eq:w projection function def}
\end{eqnarray}
and $G(l,m,w)$ is the Fourier transform of $\tilde{G}(v,u,w)$ called the
W-projection function.
Given a model of the sky brightness, the visibility on the $w = 0$ plane
can be calculated using the two dimensional Fourier transform. The
visibility measurement outside the $w = 0$ plane may then be
calculated using the convolution function $\tilde{G}(u,v,w)$. Note that representing the
visibility as a convolution and using the FFT algorithm to compute the convolution
is similar to the one-dimensional chirp z-transform algorithm.
 Calculating the image for a given
set of visibility measurements is done using iterative algorithms since there is no inverse transform. The
W-projection is a minor-major cycle algorithm that receives three
dimensional visibility measurements $V(u,v,w)$ and projects the $w$
coordinate 'out' (projection on $w=0$ plane). The 2D visibilities
$V(u,v,w=0)$ are used to calculate the reconstructed image in the
$(l,m)$ domain by a two dimensional Fourier Transform. Then a
deconvolution is performed (such as CLEAN) on the resulting image.
 The W-projection algorithm
has both high performance and high computational speed.

\subsection{The LS-MVI algorithm}
We now describe a recent approach that enables the use of modern array processing algorithms in the framework of image
deconvolution. The method will be demonstrated on simulated and measured data. However, in contrast to the CLEAN
algorithm it is in initial research stages and further development of the technique is an interesting research problem.
The LS-MVI algorithm is a novel matrix based sequential source removal algorithm originally proposed in
\cite{leshem2000a} and further improved in
\cite{bendavid08}. It is based on matrix based approach to direction-of-arrival (DOA)
estimation techniques. We would like to replace the vectors $\va_k(\vs)$ in
(\ref{eq:dirty_img_func_of_corr_in_param_form}) by a set of beamforming vectors $\vw_k(s), k=1,...,K$.
The main goal of the LS-MVI is to eliminate interference from other points in the image when estimating the
location and power of a given source. To that end, filterbank
techniques such as the MVDR and its extensions have proven very
effective. Minimizing the interference from sidelobes of the dirty
beam while observing a point source in direction $\bs=(l,m)$ can be
formulated as a constrained beamforming problem
(For simplicity we denote $\bw_k(\bs)$ by $\bw_k$ and assume that
$\vw=\left[\vw_1,...,\vw_K \right]^T$).
\begin{equation}
\bea{l}
    \vwh(\vs) \;=\; \argmin_{\vw}\; \sum_{k=1}^K \vw_k\rH \mRh_k \vw_k \\
    \qquad\hbox{\ subject to \ } \quad \\
   \sum_{k=1}^K \vw_k\rH\va_k(\vs) = 1 \,.
   \ena
\end{equation}
The solution is given by
\begin{equation}
\bea{l}
    \vwh_k(\bs) = \gb(\bs) \mRh_k^{-1} \va_k(\vs), \\
\ena
\end{equation}
where, $\gb(\bs) = \frac{1}{\sum_{k=1}^K \va_k\rH(\vs) \mRh_k^{-1} \va_k(\vs)}$,
${\bf a}_k(\vs)$ is given in equation (\ref{eq:ak vec def}) and
$\mRh_k$ is the covariance matrix measured at time $t_k$.
The vectors $\vwh(\bs)$ have different magnitudes for
different values of $\bs$. This is undesirable since it generates noise related spatial features. Therefore, the
adapted angular response (AAR) solution normalizes the norm of $\vw$ to 1.
The resulting solution is given by \beq
  \label{eq:AAR dirty image def}
  \bea{lcl}
  I_D^{\rm AAR}(l,m) & \equiv & \frac
  {\sum_{k=1}^K {\bf a}_k(l,m) ^ H \mRh_k ^{-1} {\bf a}_k(l,m)}
  {\sum_k {\bf a}_k(l,m) ^ H \mRh_k ^{-2} {\bf a}_k(l,m)}.
\ena
\eeq
This modified dirty image replaces the classical dirty image in the LS-MVI deconvolution process.

The intensity estimation used by the LS-MVI algorithm is a LS estimation of a point source at
location $(l,m)$ and given by the following equation:
\beq
\label{eq:LS MVI alpha est by LS}
\bea{ll}
\alpha &={\rm arg} \min_{\alpha}
\sum_k \| \mRh_k - \alpha {\bf
a}_k(l,m) {\bf a}_k^H(l,m)\|_F^2 \\
 & \text {subject to } \alpha \geq 0
\ena \eeq This estimate of the source power has been independently
used in ASP-CLEAN \cite{bhatnagar04}.
The closed form solution of equation (\ref{eq:LS MVI alpha est by LS}) is
given by
\beq
 \label{eq:LS MVI alpha solution}
\alpha = \max \left\{\frac{{\bf h}^H \vr}{{\bf h}^H {\bf h}},0  \right\}.
\eeq
where
\[
h\equiv \left [vec(\va_1(l,m) \va_1^H(l,m))^T,...,vec(\va_K(l,m) \va_K^H(l,m))^T\right]^T
\]
and $\vr \equiv \left[vec(\mRh_1)^T,...,vec(\mRh_K)^T\right]^T$ are obtained by stacking the array response and
the measured covariance
matrices respectively.

The intensity estimation can be improved by adding the semi-definite
constraint
\begin{equation}
\label{eq:LS-MVI improve alpha est} \mRh_k - \sigma^2 {\rm I} - \alpha
{\bf a}_k(l,m) {\bf a}_k^H(l,m) \succcurlyeq 0.
\end{equation}
The intensity estimation is bounded between the solution
(\ref{eq:LS MVI alpha solution}) and $0$. Hence, a better intensity
estimation can be achieved using a simple bi-section. A summary of the LS-MVI algorithm is given in
Table (\ref{tab:LS-MVi algo flow}).
Another improvement that has low computational complexity is to use
a joint LS estimate of all previously estimated sources. Assuming that we have collected $L$ components the
estimator is given by:
\beq \label{eq:LS MVI alpha est by_LS_cotton_schwab}
\bea{ll} \vga &={\arg} \min_{\vga}
\sum_{k=1}^K || \small{\vr_k - \sum_{i=1}^L \alpha_{i} \vq_{ki}}||^2 \\
 & \text {s.t.} \alpha_i \geq 0 {\hbox {\ for all \ } i}
\ena
\eeq
where $\vga=[\alpha_1,...,\alpha_L]$, $\vr_k=vec(\mR_k)$ and $\vq_{ki}=vec\left(\va_k(l_i,m_i) \va_k^H(l_i,m_i)\right)$.
Similarly to the CLEAN algorithm this improvement can be implemented only at major cycles, after several sources have been estimated.
\begin{table}
  \begin{tabular}{|p{0.45\textwidth}|}
  \hline
  \textbf{Initialization}: \\
  \quad $\bullet$ $\mR_k^0 = \mR_k$, $\forall k = 1,\ldots,K$\\
  \quad $\bullet$ Calculate $I_D^{\rm AAR}$ using eq. (\ref{eq:AAR dirty image def})\\
  \quad $\bullet$ $i = 0$ \\
   \quad $\bullet$ $B_{\rm rec} = {\rm Gaussian}$ \\
  \textbf{While $I_D$ is not noise like}:\\
  \quad $\bullet$ $(l_i,m_i) = {\rm arg} \max I_D^{\rm AAR}(l,m)$ \\
  \quad $\bullet$ Estimate $\alpha_i$ according to eq.(\ref{eq:LS MVI alpha solution}) \\
 \quad $\bullet$ Optionally improve $\alpha_i$ estimation
 according to eq. (\ref{eq:LS-MVI improve alpha est}) \\
  \quad $\bullet$ $\mR_k^{i+1} = \mR_k^i - \gamma \alpha_i {\bf
  a}_k(l_i,m_i) {\bf a}_k^H(l_i,m_i)$, $\forall k = 1\ldots K$\\
   \quad $\bullet$ Calculate $I_D^{\rm AAR}$ using $\mR_k^{i+1}$ using eq. (\ref{eq:AAR dirty image def})\\
   \quad $\bullet$ $ i = i + 1$ \\
 \textbf{Finalize}:\\
  \quad $\bullet$ $I_{\rm rec} = \sum_i \gamma \alpha_i B_{\rm rec}(l - l_i, m -
  m_i)$ \\
   \hline
\end{tabular}
  \caption{LS-MVI algorithm flow}
  \label{tab:LS-MVi algo flow}
\end{table}

There are two main differences between LS-MVI and CLEAN. First, the
LS-MVI uses a different type of dirty image and second, the LS-MVI
performs a more sophisticated intensity estimation than CLEAN. The
dirty image used by the LS-MVI is  $I_D^{\rm AAR}$ given in equation
(\ref{eq:AAR dirty image def}). The main advantage of
the AAR dirty image over simple MVDR is the isotropic noise response
that prevents the formation of spatially varying noise related
artifacts. In \cite{bendavid08} further extensions for enforcing
semi-definite constraints in a Cotton-Schwab type of iteration are
also presented. It should also be noted that there is no need to
compute the complete dirty image in order to find the maximum and
optimization techniques can do this much faster, especially if the
user can provide windows similar to CLEAN windows currently used by
radio astronomers. Like CLEAN, the LS-MVI should be implemented in
the visibility domain to eliminate gridding effects.
\section{Global optimization based techniques}
We now turn to a second family of solutions to the image formation problem. These
solutions are based on optimizing a global property of the image subject to goodness of fit to
the data. They vary from least squares (LS) based techniques to maximum entropy and
$\ell_1$ based reconstruction.
\subsection{Linear deconvolution}
Computationally the simplest way to solve the image formation
problem is through linear inversion. There are two main approaches
in this area: The well known Least Squares (LS) technique and Linear
Minimum Mean Square Error (LMMSE). Such techniques can work well
when the $(u,v)$ coverage is good and the inversion is well
conditioned. Furthermore linear inversion can work independently of
the complexity of the source structure. However, linear techniques
can result in significant noise enhancement in ill-posed problems.
For a fully sampled visibility domain, these techniques can provide
a first approximation to the image. To overcome this problem one can
use a constrained LS, also known as non-negative LS (NNLS), first
proposed for radio synthesis imaging by Briggs \cite{briggs95}. The
idea is that the image is positive. Putting these constraints into
the deconvolution, yields a computationally expensive, though
feasible algorithm. An excellent overview of the implementation of
the NNLS can be found in \cite{briggs95}.

\subsection{Maximum Entropy image reconstruction}
The maximum entropy image formation technique is one of the two most popular deconvolution techniques
in radio astronomy (together with CLEAN).
The maximum entropy principle was first proposed by Jaynes \cite{jaynes57}. A good
overview of the philosophy behind the idea can be found in \cite{jaynes82}. Since then it has been used in a wide
spectrum of imaging problems.
The basic idea behind MEM is the following:  Out of all the images which are consistent with the measured
data where the noise distribution does not satisfy the positivity demand,i.e., the sky brightness is a positive function, consider only those that satisfy the positivity demand. From these select the one that is most likely to have been
created randomly. This idea was also proposed by \cite{frieden72} for optical images and applied to
radio astronomical imaging in \cite{gull78}. Other approaches based
on differential entropy have also been suggested \cite{ables74} and
\cite{wernecke77}. An extensive collection of papers discussing these different
methods and aspects of maximum entropy can be found in a number of papers in
\cite{roberts84}. \cite{narayan86} provides an overview of various maximum entropy techniques and the use of the
various options for choosing the entropy measure. Interestingly, in that paper, a closed form solution is given for
the noiseless case, but not for the general case.

The approach in \cite{gull78} begins with a prior image and iterates
between maximizing the entropy function and updating the $\chi^2$ fit to the data. The computation of the image based on
a prior image is done analytically, but at each step the model visibilities are updated,
through a two-dimensional Fourier transform.
This type of algorithm is known as a fixed point algorithm, since it is based on iterating a function until it converges
to a fixed point.

The maximum entropy solution is given by solving the
following Lagrangian optimization problem \cite{gull78}:
\beq
\label{opt_mem}
I^{MEM}=\arg\max_{I}-\sum_{l,m} I(l,m) \log \frac{I(l,m)}{F(l,m)}-\frac{\gl}{2}\chi^2(V),
\eeq
where
\beq
\chi^2(V)=\sum_{(u,v) \in A} \frac{1}{\gs^2}\left|\Vhat(u,v)-
V(u,v) \right|^2,
\eeq
$V(u,v)$ are the model based visibilities, $\gl$ is a Lagrange multiplier for the constraint that
$V(u,v)$ should match the measured visibilities $\Vhat(u,v)$,
$A$ is the $(u,v)$ coverage of the radio telescope and $F(l,m)$ is a reference image.
Taking the derivative with respect to $I(l,m)$ we obtain that the solution is given by:
\beq
\label{update_I}
I(l,m)=exp\left(-1+\log F(l,m)+\gl \Delta(l,m)\right)
\eeq
where
\[
\Delta(l,m)=\sum_{(u,v)\in A} \frac{1}{\gs^2}Re\left(\left(\Vhat(u,v)-
V(u,v)\right) e^{\frac{\jmath 2\pi (ul+vm)}{N}}\right).
\]
The basic maximum entropy algorithm now proceeds by choosing an initial image model
(typically a flat image or a low resolution image) computing the model based visibilities
$V(u,v)$ on a grid $A$. Using these visibilities a new model
image is computed by equation (\ref{update_I}). New visibilities are computed from the new model and the process is
iterated until convergence.

While it is known that for the maximum entropy, this approach usually converges,
the convergence can be slow \cite{narayan86}. Improved methods based on the Newton method and the Conjugate
Gradient technique were put forward by \cite{cornwell85, sault90, skilling84}. These methods perform direct optimization of the entropy
function subject to the $\chi^2$ constraint. Generalization of the maximum entropy using wavelets and multi-resolution techniques have also been proposed (see
e.g., \cite{pantin1996}, \cite{maisinger2004}).

\subsection{Compressed sensing and sparse reconstruction techniques}
Recently there has been growing interest in using $\ell_1$ based cost functions for deconvolution (see \cite{levanda08} and \cite{wiaux09} and unpublished notes by Schwardt).
This renewed interest in $\ell_1$ comes from
recent results related to compressed sampling using Fourier bases.
It is worth noting that as early as 1987 Marsh and Richardson \cite{Marsh1987} proved that the CLEAN algorithm can be
regarded as an $\ell_1$ minimization for a single point
source image. $\ell_1$ is not the only criterion. Recovery of noisy and
blurred images using total variation ($TV$) optimization
for smooth images was discussed by Dobson and Santosa
\cite{Dobson1996}. Chen et al. \cite{Chen1998} dealt with $\ell_1$ minimization
of an image basis to achieve image sparseness using linear programming.
Feuer and Nemirovski \cite{Feuer2003} and Elad and Bruckstein \cite{Elad} established sufficient and necessary
conditions for replacing $\ell_0$ optimization (computing the sparsest solution with high
computational complexity) by linear programming when searching for
the unique sparse representation.  Rudelson and Vershynim \cite{Rudelson2006} proved the  best
known guarantees for exact reconstruction of a sparse signal from its Fourier measurements.

Radio astronomical image reconstruction is done based on the visibility measurement in the $(u,v)$ domain.
Reconstruction of the source image $I(l,m)$ is equivalent to
estimating the missing visibility points. The missing $V(u,v)$
measurements together with the image itself are estimated by
minimizing a cost function $||I(l,m) ||_{\ell_1}$ in the
$(l,m)$ domain using the constraints of image positivity and the measured visibility data. Note that
since $I(l,m)$ is a positive quantity we have:
\begin{equation}
||I(l,m) ||_{\ell_1} = \sum_{l = 1}^{N} \sum_{m = 1}^{N} I(l,m)
\end{equation}
which allows us to use linear programming.
To solve the reconstruction
problem fast, we represent the problem as a linear programming
problem with real variables. To that end let $\langle\cdot,\cdot
\rangle$ be a one-to-one pairing function mapping $\left\{0,...,N-1
\right\}\times \left\{0,...,N-1 \right\}$ onto
$\left\{0,...,N^2-1\right\}$. Let $\mF$ be an $N^2\times N^2$ matrix
whose elements satisfy
\beq \mF_{\langle l,m\rangle,\langle u,v
\rangle}=e^{\frac{-2\pi \jmath}{N}\left(ul+vm\right)}.
 \eeq
 Let $\vgx=\hbox{vec}(\mV)$ and let $\vt=\hbox{vec}(\mI)$. We have
\beq \label{Complex-Fourier} \vgx = \mF \vt . \eeq Note that $\vt$
is a real vector since the visibility measurements satisfy
$V(u,v)={\bar{V}(-u,-v)}$. To make the problem real we define
$\mF_R=\Re(\mF),\mF_I=\Im(\mF)$ and variables
$\vgx_R=\Re(\vgx),\vgx_I=\Im(\vgx)$. (\ref{Complex-Fourier}) now
becomes
\begin{eqnarray}
\label{real_Fourier}
 \vgx_R &=& \mF_R  \vt \\ \nonumber
  \vgx_I &=& \mF_I  \vt \\ \nonumber
 \end{eqnarray}
For the measured locations $(u_i,v_i)$ we have:
\begin{equation}
 \label{constrain measured visibilities}
 \bea{c}
\vgx_R(\langle u_i,v_i\rangle)=\Re\left(\Vhat(u_i,v_i)\right) \quad i=1,...,M \\
\vgx_I(\langle u_i,v_i\rangle)=\Im\left(\Vhat(u_i,v_i)\right) \quad i=1,...,M \\
\ena
\end{equation}
where $M$ is the number of given measurements in the $(u,v)$ domain.
The linear programming problem is described in Table
\ref{linear_prog} (for more details the reader is referred to
\cite{levanda08}).
\begin{table} [htb]
\centering
\begin{tabular}{|l|}
\hline
$\min_{{\vt} }\sum_{i = 1}^{N^2}{t_i}$\\
Subject to \\
\quad $\vgx_R(\langle u_i,v_i\rangle)=\Re\left(\Vhat(u_i,v_i)\right)$\\
\quad $\vgx_I(\langle u_i,v_i\rangle)=\Im\left(\Vhat(u_i,v_i)\right)$\\
\quad $\vzero \leq \vt$\\
\hline
\end{tabular}
\caption{$\ell_1$ optimization using linear programming}
\label{linear_prog}
\end{table}
In \cite{wiaux09} a joint $\ell_1$ and $\ell_2$ is also discussed. This makes it possible to include prior knowledge on the
noise power. Using the total variation is also a possibility that
leads to $\ell_1$ optimization. Note that using total variation and maximum entropy are related since both
functionals impose smoothness on the image.
\section{Self calibration and robust MVDR for synthetic aperture arrays}
We now turn to the case where the array response is not completely known, but we have some statistical
knowledge of the error, e.g., we know the covariance matrix of the array response error at each epoch (measurement time).
Typically this covariance will be time invariant or will have slow temporal variation.
In this case we extend the robust dirty image as described in \cite{leshem2004} to the synthetic aperture array case.
This generalization follows the analysis in \cite{bendavid08}.
Since the positive definite constraint on the residual covariance matrices is important in our application
we extended the robust Capon estimator of \cite{stoica03}.
To that end assume that at each epoch we have an uncertainty ellipsoid describing the uncertainty of the
array response (as well as unknown atmospheric attenuation).
This is described by
\beq
\left(\va_k(\vs)-\vab_k(\vs) \right)^H \mC_k \left(\va_k(\vs)-\vab_k(\vs) \right) \le 1
\eeq
where $\vab_k(\vs)$ is the nominal value of the array response towards the point $\vs$
and $\mC_k$ are the covariance matrices of the uncertainty in the calibration parameters at time $k$.
Generalizing the LS-MVI we would like to solve the following problem:
\beq
\label{robust_capon_moving}
\bea{ll}
\left[\grh,\vah_1,...,\vah_k \right]=\arg \max_{\gr,\va_1,...,\va_k} \gr & \\
\hbox{subject to} & \\
\mRh_k- \gs^2 \mI-\gr \va_k \va_k^H\succeq \vzero & k=1,...,K  \\
\left(\va_k(\vs)-\vab_k(\vs) \right)^H \mC_k \left(\va_k(\vs)-\vab_k(\vs) \right) \le 1 & k=1,...,K
\ena
\eeq
Let $\gt=1/\gr$.
The problem (\ref{robust_capon_moving}) is equivalent to the following problem
\beq
\bea{ll}
\left[\hat \gt,\vah_1,...,\vah_k \right]=\arg \min_{\gt,\va_1,...,\va_k} \gt & \\
\hbox{subject to} & \\
\left[
\bea{cc}
\mRh_k- \gs^2 \mI &  \va_k \\
\va_k^H           & \gt
\ena
\right]\succeq \vzero & k=1,...,K \\
\left[
\bea{cc}
\mC_k & \left(\va_k(\vs)-\vab_k(\vs) \right) \\
\left(\va_k(\vs)-\vab_k(\vs) \right)^H          & 1 \ena
\right]\succeq \vzero & k=1,...,K. \ena \eeq This problem is once
again a semi-definite programming problem that can be solved
efficiently via interior point techniques \cite{boyd96}. We can now
replace the MVDR estimator by this robust version. Interestingly, we
obtain estimates of the corrected array response $\vah(\vs_k)$.
Using the model we obtain for each $k$ \beq \va_k(\vs)=\mgG_k
\vab_k(\vs) \eeq Hence, the self-calibration coefficients can be
estimated using least squares fitting \beq \hat \mgG_k=\arg
\min_{\grg_1,...,\grg_p} \sum_{\ell=1}^L
\|\vah_k(\vs_\ell)-\mgG_k\vab_k(\vs_\ell)\|^2 \eeq where
$\mgG_k=\diag\{\grg_{k,1},...,\grg_{k,p}\}$. Of course, when the
self-calibration parameters vary slowly we can combine the
estimation over multiple epochs. This might prove instrumental in
calibration of LOFAR type arrays, where the calibration coefficients
vary across the sky. Since the computational complexity of the self-calibration semi-definite programming is higher than that of the
MVDR dirty image, it is too complicated to solve this problem for
each source in the image. Hence it should be used in a way similar to the
external self-calibration cycle (\cite{pearson84}), where this problem is solved using
a nominal source locations model. The advantage over ordinary self-calibration is that beyond the re-evaluation of the calibration
parameters, we obtain better estimates of the source powers,
without significant increase in the complexity. Another interesting
alternative is to use the doubly constrained robust Capon beamformer which combines a
norm constraint as in the AAR dirty image with robust Capon
beamforming \cite{li2004}. 
\section{Examples and Comparisons}
In this section we describe three examples of the various algorithms,
including a simulated example of an extended source, an example from
the LOFAR test station and an example of Abell 2256 observed by the
VLA\footnote{Initial calibration was done by Tracy Clarke}.
\subsection{Simulated Extended Source}
\begin{figure}
\includegraphics[width=0.4\textwidth]{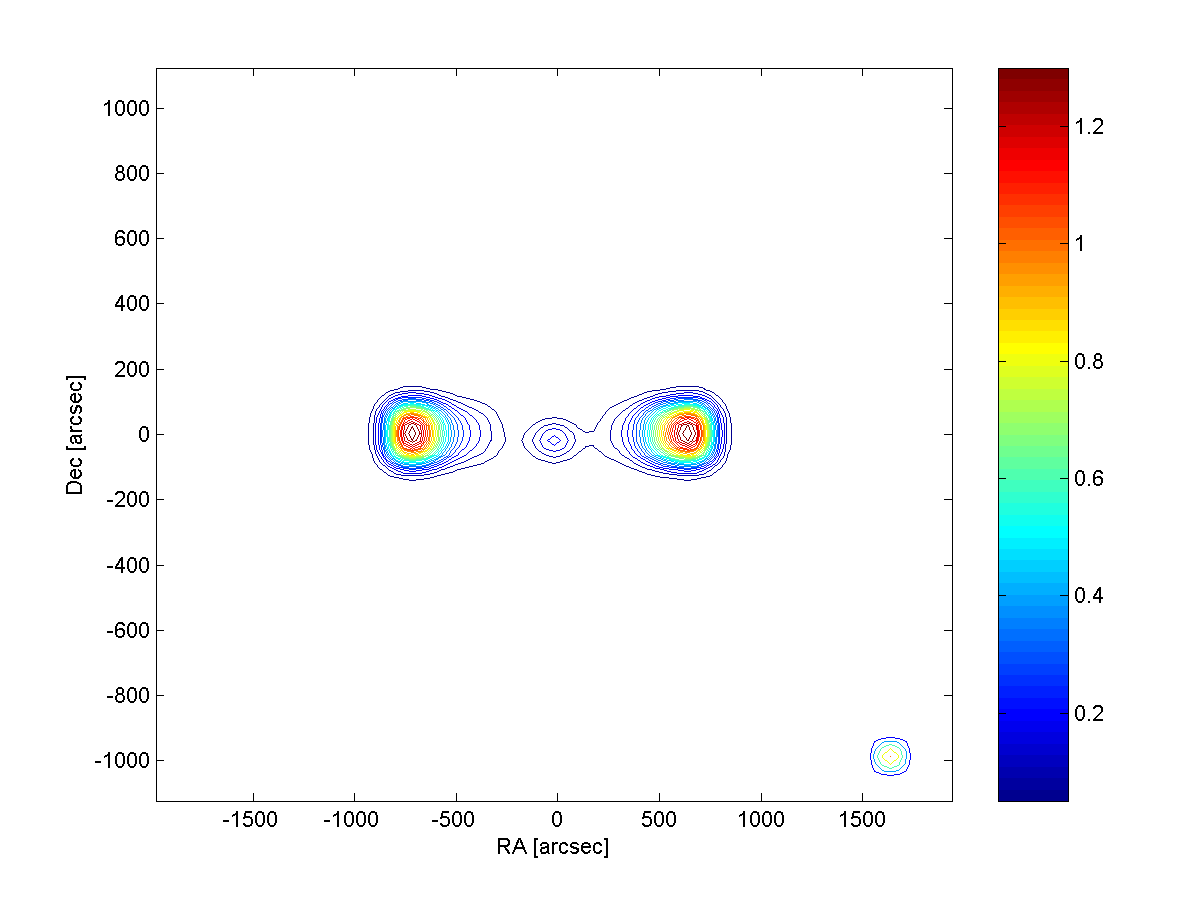}
\caption{Original extended source image}
 \label{fig:exp sim orig src}
\end{figure}

\begin{figure}
\begin{tabular}{|cc|}
\hline
  &   \\
\includegraphics[width=0.2\textwidth]{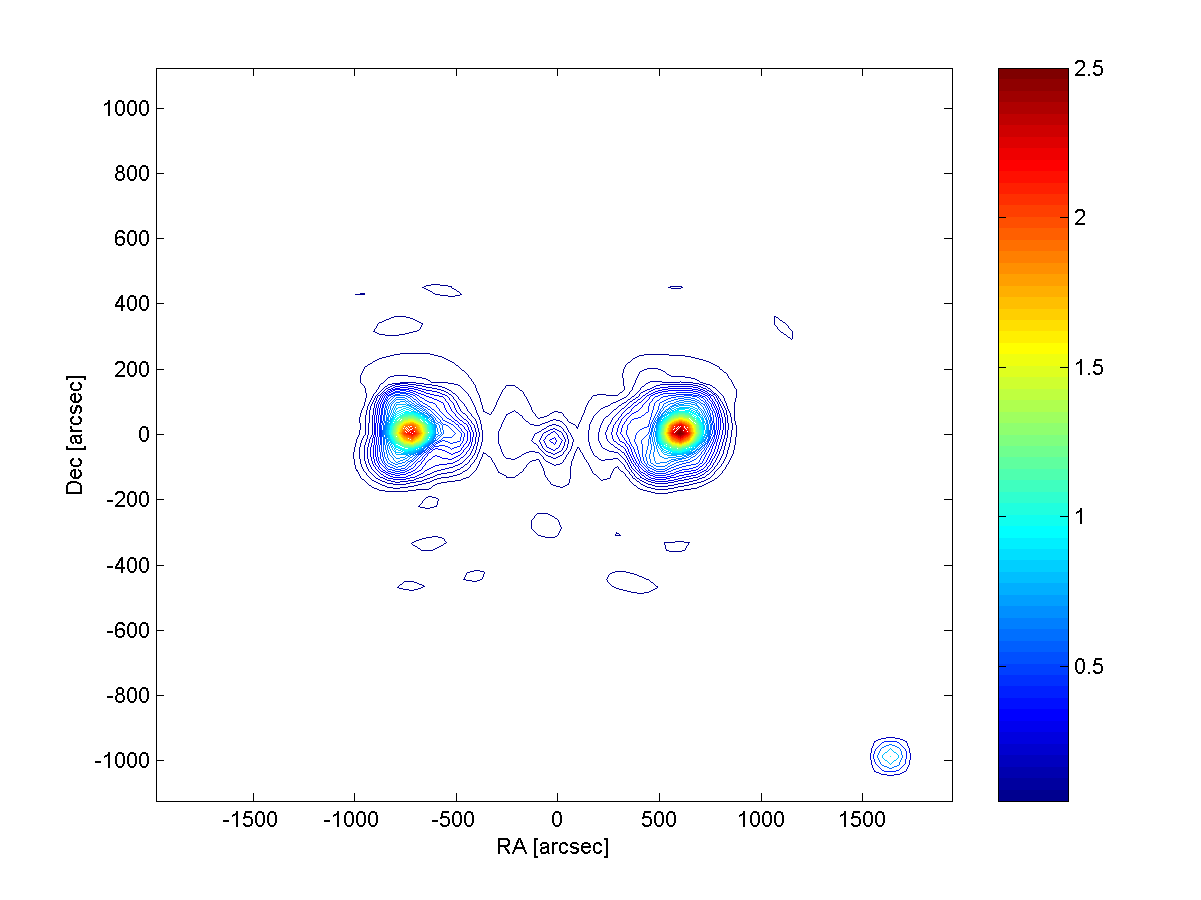} & \includegraphics[width=0.2\textwidth]{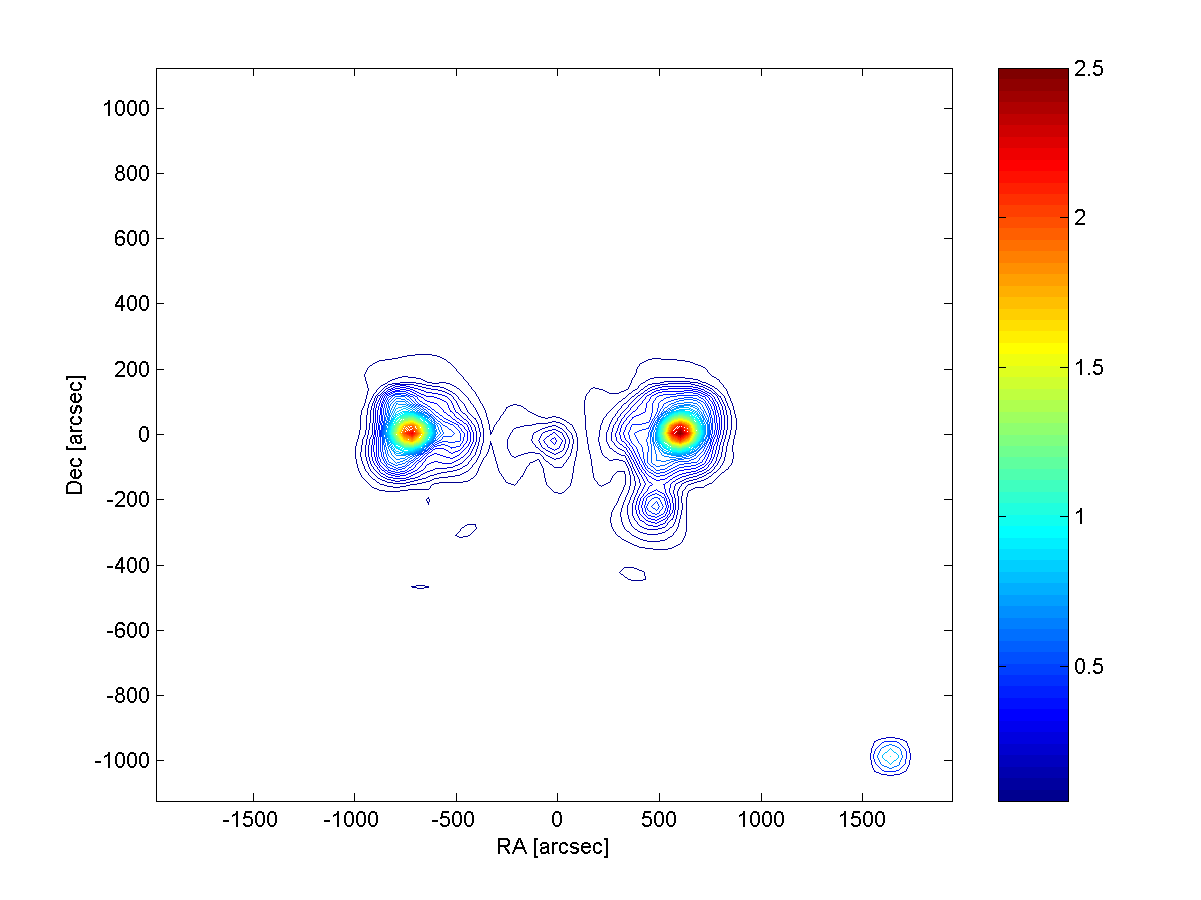} \\
{\tiny (a) } & {\tiny (b) } \\
 \includegraphics[width=0.2\textwidth]{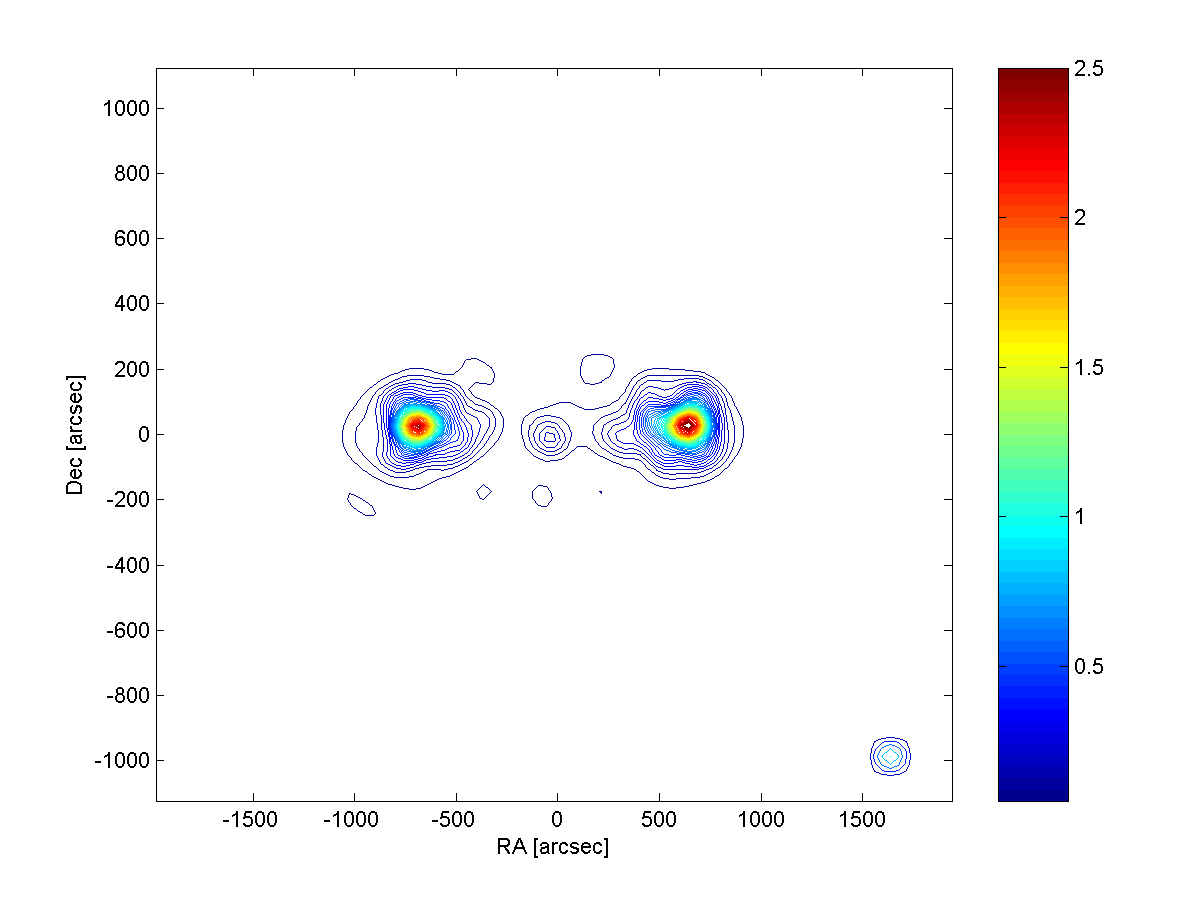} & \includegraphics[width=0.2\textwidth]{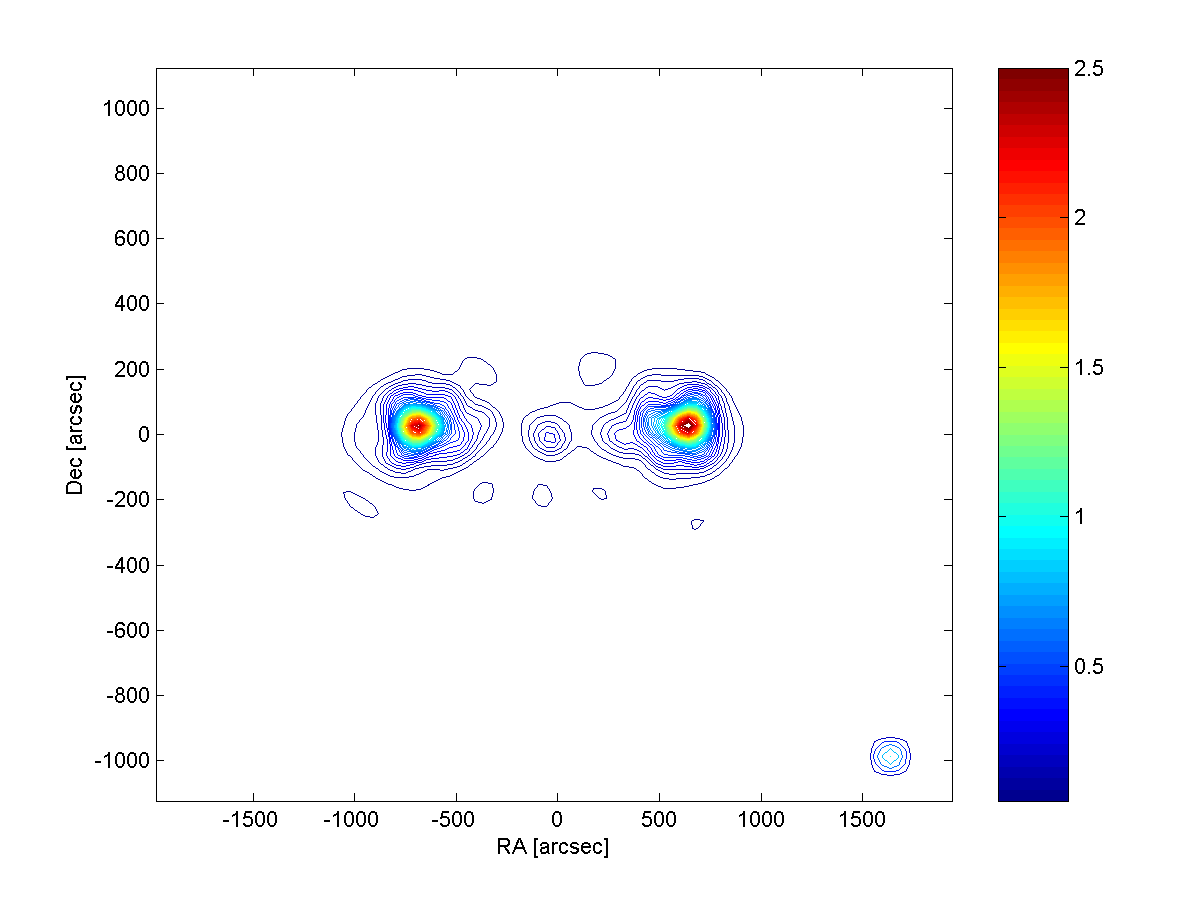} \\
 {\tiny (c) } & {\tiny (d) } \\

\hline
\end{tabular}
\caption{Reconstructed images of the CLEAN and LS-MVI algorithms.
(a) CLEAN reconstructed image after 100 iterations. (b) CLEAN
reconstructed image after 120 iterations. (c) LS-MVI reconstructed
image after 100 iterations. (d) LS-MVI reconstructed image afer 300
iterations.} \label{fig:exp sim clean and ls mvi}
\end{figure}

An extended source (Figure (\ref{fig:exp sim orig src})) was
simulated using an East-West array containing 10 antennas
logarithmically spaced up to 1000$\lambda$.
 CLEAN deconvloution results are depicted in Figure
(\ref{fig:exp sim clean and ls mvi} a) and (\ref{fig:exp sim clean
and ls mvi} b). After 100 CLEAN iterations, the center of the source
is partially reconstructed with distortion. After additional 20
iterations an artifact is generated (below the strong point on the
right). This divergence can often occur in CLEAN when applying it to
extended sources. The LS-MVI results are presented in Figure
(\ref{fig:exp sim clean and ls mvi} c) and (\ref{fig:exp sim clean
and ls mvi} d). After 100 iterations the center of the source is
reconstructed and after 200 additional iterations and reconstruction
is stable. The reason for this is the fact that CLEAN overestimate the power due to the high sidelobes level.
Further analysis of this example is given in \cite{bendavid08}.

%
\subsection{LOFAR Test Station Data}
\begin{figure}
\begin{tabular}{|cc|}
\hline
  &   \\
\includegraphics[width=0.2\textwidth]{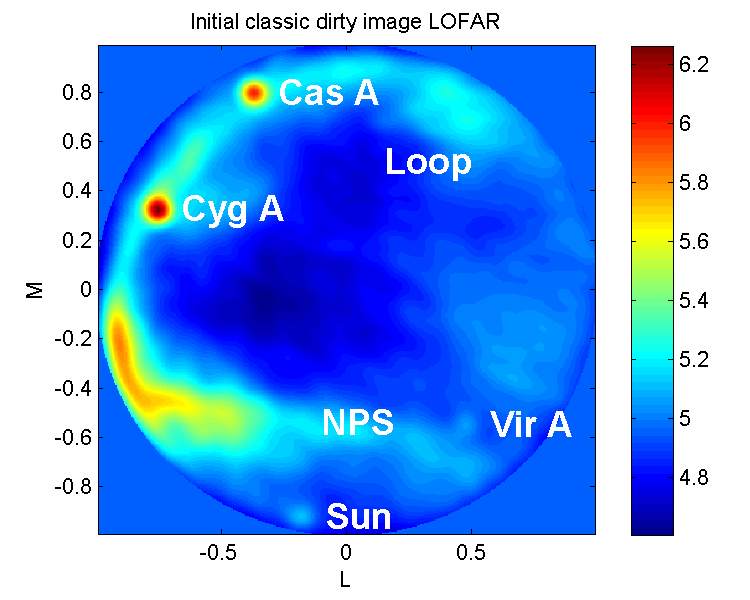} & \includegraphics[width=0.2\textwidth]{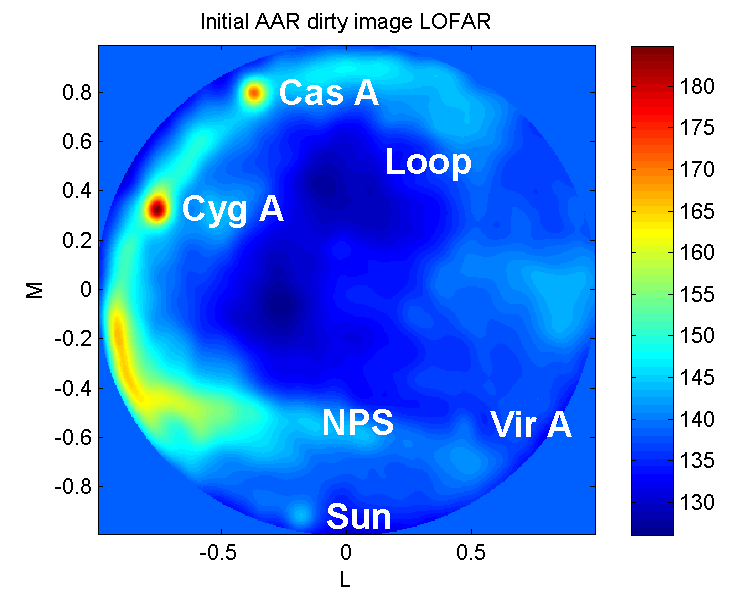} \\
{\tiny (a) } & {\tiny (b) } \\
\includegraphics[width=0.2\textwidth]{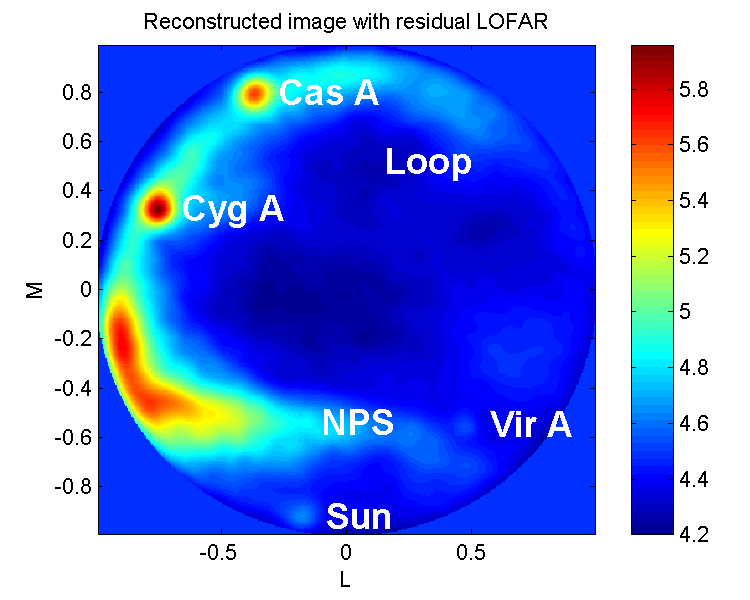}
&
\includegraphics[width=0.2\textwidth]{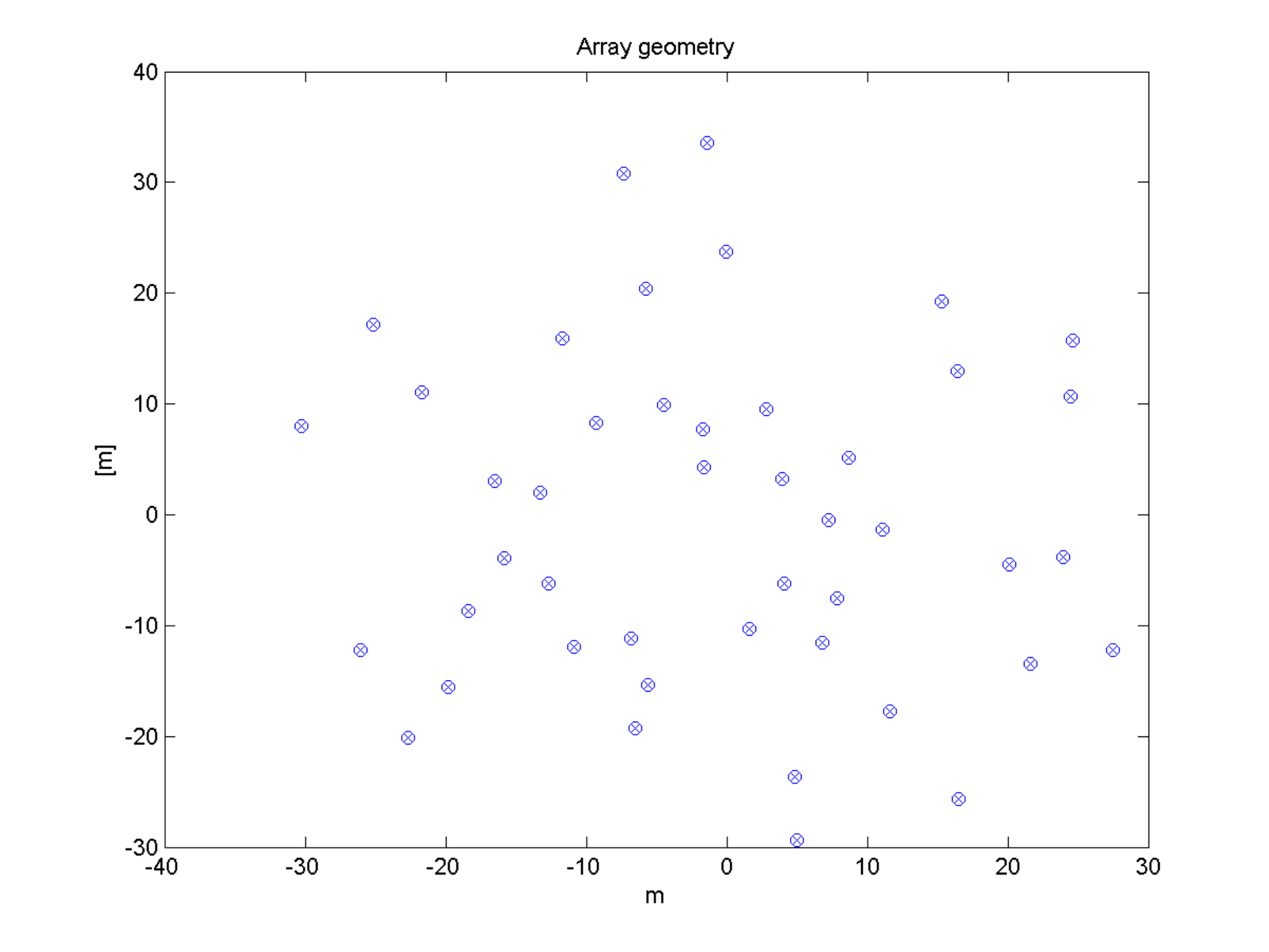}
\\
{\tiny (c) } & {\tiny (d) } \\
\hline
\end{tabular}
\caption{LOFAR station  images. (a) classic dirty image. (b) AAR
dirty image. (c) Reconstructed image using AAR based cleaning. (d)
Array geometry} \label{fig:exp LOFAR  images}
\end{figure}

The LOFAR test station data were recorded using $25$ frequency bands
of 156kHz using 45 antennas (array geometry is given in Figure
(\ref{fig:exp LOFAR images} d)). The data were calibrated by
S. Wijnholds. The AAR dirty image and the classic dirty image are given
in Figure (\ref{fig:exp LOFAR  images}a) and (\ref{fig:exp LOFAR
images}b), respectively. Since the LOFAR station benefits from a
good $(u,v)$ domain coverage, the two dirty images are similar. The
reconstructed image using the LS-MVI algorithm is displayed in
Figure (\ref{fig:exp LOFAR images} c); the spurious emission on the
right side of the image was removed.
\subsection{Abell  2256}
\begin{figure}
\begin{tabular}{|cc|}
\hline
  &   \\
  \includegraphics[width=0.2\textwidth]{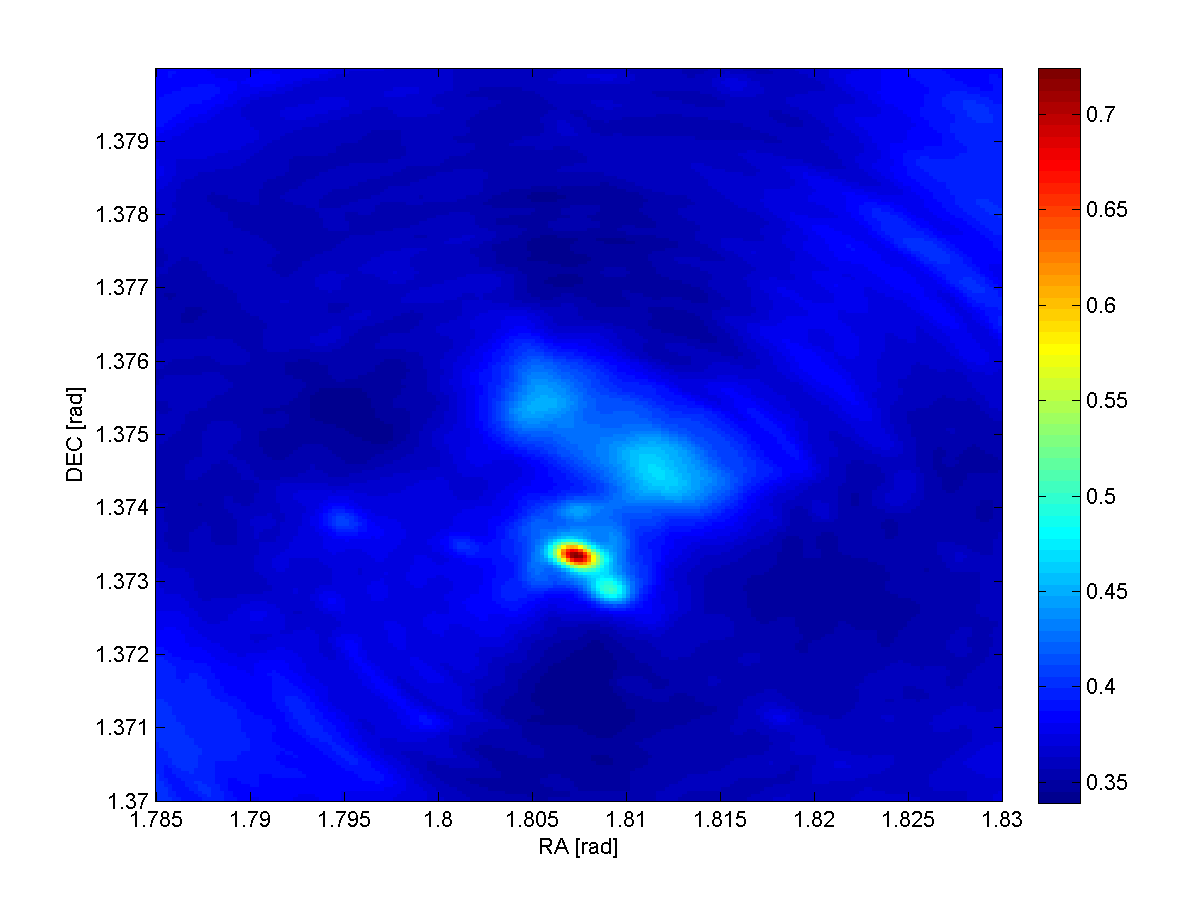} &
 \includegraphics[width=0.2\textwidth]{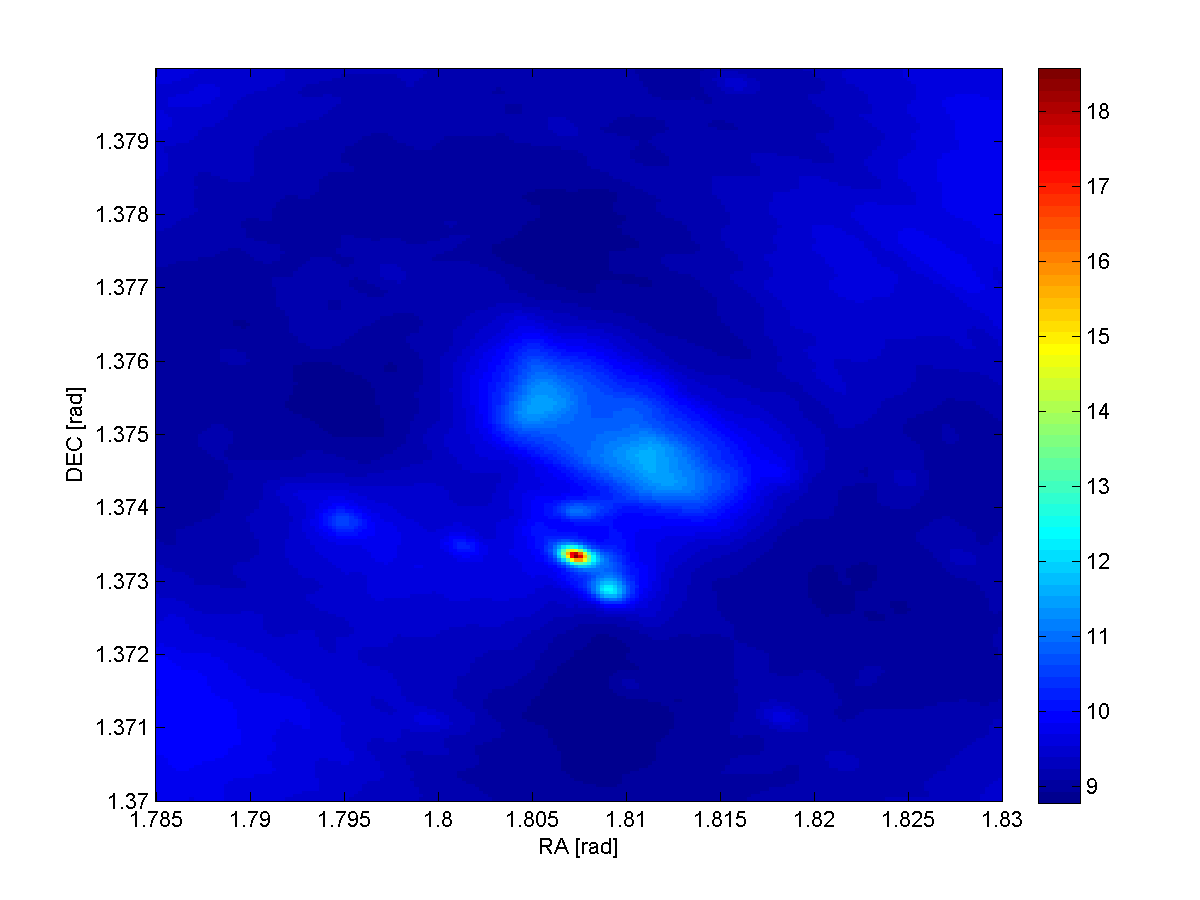} \\
{\tiny (a) } & {\tiny (b) } \\
\includegraphics[width=0.2\textwidth]{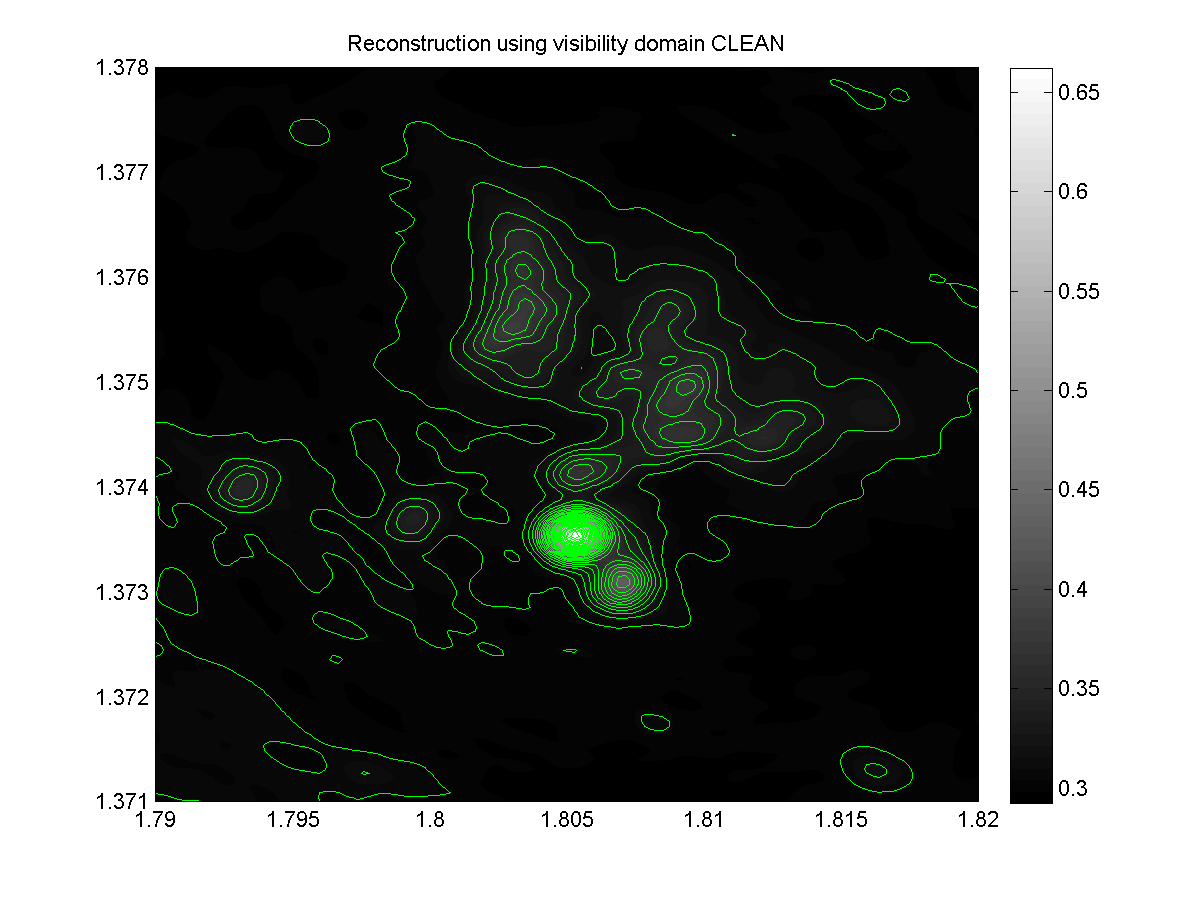} &
 \includegraphics[width=0.2\textwidth]{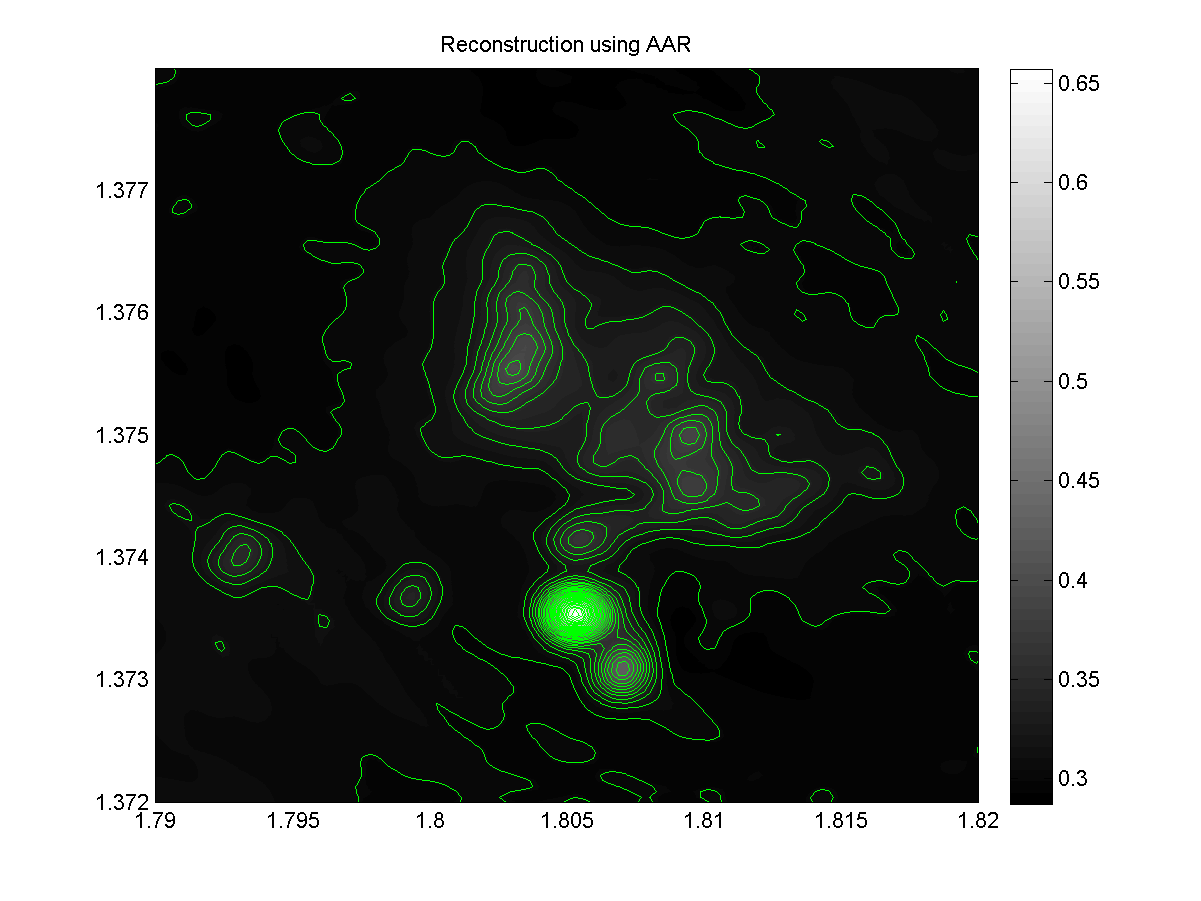} \\
{\tiny (c) } & {\tiny (d) } \\
\hline
\end{tabular}
\caption{Abell 2256  images. (a) Initial classic dirty image. (b) Initial AAR dirty image. (c) classic CLEAN
reconstructed image. (d) LS-MVI reconstructed image.}
\label{fig:exp Abell  images}
\end{figure}

The last example used VLA data of Abell 2256 \footnote{Abell 2256 is a merging of two (or three)
large clusters of more than 500 galaxies. It exhibits strong radio emissions and is one of the strongest X-ray emitters
\cite{rottgering1994}.} The data measured by Clarke and Ensslin \cite{clarke2006} contain a single frequency
band around 1369 MHz. The data were processed using both CLEAN and LS-MVI algorithms for 30 iterations
\footnote{This is a first example of application of the LS-MVI algorithm for measured data. As such it is only
a preliminary example and significant improvements can be made, e.g., in \cite{clarke2006} data were also
self calibrated using phase data and then amplitude and phase. This is required here in order to achieve a deeper level of cleaning.}.
We used a visibility domain CLEAN (updates were performed on the ungridded visibility).
The initial data (dirty image) of the CLEAN are shown in Figure (\ref{fig:exp Abell  images} a),
The strong sidelobes structure is clearly visible as large circles in the dirty image. In contrast the
initial AAR dirty image is shown in Figure (\ref{fig:exp Abell  images} b). The sidelobes level is much
lower and several point sources that are invisible in the classical dirty image are now visible.
The reconstruction using the visibility domain CLEAN  is shown in Figure (\ref{fig:exp Abell  images} c). The
sidelobes level is reduced and the source structure is clearly seen. The reconstruction using
the LS-MVI algorithm is shown in Figure (\ref{fig:exp Abell  images} d). Similar to the CLEAN, the sources structure
is visible and the sidelobes level is significantly reduced. It should be emphasized that even though we have used only 30
iterations, the strong structure is consistent with that of \cite{clarke2006} and \cite{bridle76}.

\section*{Acknowledgements}
We would like to thank T. Clarke, H. Intema, H. Rottgering and S. Wijnholds  for providing the data used to demonstrate
the various techniques, to Seth Shostak and the SETI institute for providing the photo of the Allen Telescope Array
and NRAO for permission to use VLA images. We would also like to thank the anonymous reviewers and the guest editor
A-J. van der Veen for comments that significantly enhanced the presentation.
\section*{Authors}
\bds
\item {\em Ronny Levanda} received her B.Sc. in Physics and M.Sc. in neural networks, from the Tel Aviv
University, in 1995 and 2000 respectively. She has 10 years of algorithm development experience, in High-Tec
companies. She is now studying towards her Ph.D. in Bar-Ilan University.
\item {\em Amir Leshem}
Amir Leshem received the B.Sc. degree (cum laude) in mathematics and physics, the M.Sc. degree (cum laude)
in mathematics, and the Ph.D. degree in mathematics, all from the Hebrew University, Jerusalem, Israel.
He is one of the founders of the School of Electrical and Computer Engineering, Bar-Ilan University, Ramat Gan,
Israel, where he is currently an Associate Professor and Head of the signal processing track.
His main research interests include multichannel communication, applications of game theory to communication, array and statistical
signal processing with applications to sensor arrays and networks,
wireless communications, radio-astronomy,  and brain research, set theory, logic, and foundations of mathematics.
\eds

\begin{thebibliography}{10}

\bibitem{jansky33}
K.~Jansky, ``Electrical disturbances apparently of extraterrestrial origin,''
  {\em Proceedings of the IRE}, vol.~21, pp.~1387--1398, Oct. 1933.

\bibitem{reber1940}
G.~Reber, ``Cosmic statics,'' {\em Proceedings of the IRE}, vol.~28,
  pp.~68--70, Feb. 1940.

\bibitem{ryle1946}
M.~Ryle and D.~Vonberg, ``Solar radiation on 175 {M}c./s,'' {\em Nature},
  vol.~158, pp.~339--340, Sept. 1946.

\bibitem{penzias1965}
A.~A. {Penzias} and R.~W. {Wilson}, ``{A Measurement of Excess Antenna
  Temperature at 4080 {Mc/s}.},'' {\em Astrophysical Journal}, vol.~142,
  pp.~419--421, July 1965.

\bibitem{smoot1992}
F.~G. {Smoot} and {et al.}, ``{Structure in the COBE differential microwave
  radiometer first-year maps},'' {\em Astrophysical Journal Letters}, vol.~396,
  pp.~L1--L5, Sept. 1992.

\bibitem{hewish1968}
A.~{Hewish}, S.~J. {Bell}, J.~D.~H. {Pilkington}, P.~F. {Scott}, and R.~A.
  {Collins}, ``Observation of a rapidly pulsating radio source,'' {\em Nature},
  vol.~217, pp.~709--713, Feb. 1968.

\bibitem{ewen1951}
H.~Ewen and E.~Purcell, ``Observation of a line in the galactic radio
  spectrum,'' {\em Nature}, vol.~168, p.~356, Feb. 1951.

\bibitem{perley84}
R.~A. {Perley}, J.~W. {Dreher}, and J.~J. {Cowan}, ``{The jet and filaments in
  Cygnus A},'' {\em Astrophysical Journal Letters}, vol.~285, pp.~L35--L38,
  Oct. 1984.

\bibitem{ryle62}
M.~Ryle, ``The new {C}ambridge radio telescope,'' {\em Nature}, vol.~194,
  pp.~517--518, 1962.

\bibitem{hogbom74}
J.~A. H\"{o}gbom, ``Aperture synthesis with nonregular distribution of
  intereferometer baselines,'' {\em Astron. Astrophys. Suppl}, vol.~15,
  pp.~417--426, 1974.

\bibitem{frieden72}
B.~Frieden, ``Restoring with maximum likelihood and maximum entropy,'' {\em
  Journal of the Optical Society of America}, vol.~62, pp.~511--518, 1972.

\bibitem{gull78}
S.~Gull and G.~Daniell, ``Image reconstruction from incomplete and noisy
  data,'' {\em Nature}, vol.~272, pp.~686--690, 1978.

\bibitem{ables74}
J.~Ables, ``Maximum entropy spectreal analysis,'' {\em AAS}, vol.~15,
  pp.~383--393, 1974.

\bibitem{wernecke77}
S.~Wernecke, ``Two dimensional maximum entropy reconstruction of radio
  brightness,'' {\em Radio Science}, vol.~12, pp.~831--844, 1977.

\bibitem{cornwell85}
T.~Cornwell and K.~Evans, ``A simple maximum entropy deconvolution algorithm,''
  {\em Astronomy and Astrophysics}, vol.~143, pp.~77--83, 1985.

\bibitem{rao2009}
U.~Rau, S.~Bhatnagar, M.~Voronkov, and T.~Cornwell, ``Advances in calibration
  and imaging techniques in radio interferometry,'' {\em Proceeding of the
  IEEE}, vol.~97, pp.~1472--1481, Aug 2009.

\bibitem{pantin1996}
E.~{Pantin} and J.-L. {Starck}, ``{Deconvolution of astronomical images using
  the multiscale maximum entropy method.},'' {\em Astronomy and Astrophysics
  Supplements}, vol.~118, pp.~575--585, Sept. 1996.

\bibitem{briggs95}
D.~S. Briggs, {\em High fidelity deconvolution of moderately resolved sources}.
\newblock PhD thesis, The new Mexico Institute of Mining and Technology,
  Socorro, New Mexico, 1995.

\bibitem{leshem2000a}
A.~Leshem and A.~van~der Veen, ``Radio-astronomical imaging in the presence of
  strong radio interference,'' {\em IEEE Trans. on Information Theory, Special
  issue on information theoretic imaging}, pp.~1730--1747, August 2000.

\bibitem{bendavid08}
C.~Ben-David and A.~Leshem, ``Parametric high resolution techniques for radio
  astronomical imaging,'' {\em Selected Topics in Signal Processing, IEEE
  Journal of}, vol.~2, pp.~670--684, Oct. 2008.

\bibitem{levanda08}
R.~Levanda and A.~Leshem, ``Radio astronomical image formation using sparse
  reconstruction techniques,'' {\em Electrical and Electronics Engineers in
  Israel, 2008. IEEEI 2008. IEEE 25th Convention of}, pp.~716--720, Dec. 2008.

\bibitem{wiaux09}
Y.~Wiaux, L.~Jacques, G.~Puy, A.~Scaife, and P.~Vandergheynst, ``Compressed
  sensing imaging techniques for radio interferometry,'' {\em Monthly Notices
  of The Royal Astonomical Society}, Submitted 2009.

\bibitem{reid2006}
R.~Reid, ``Smear fitting: a new image-deconvolution method for interferometric
  data,'' {\em Monthly Notices of the Royal Astronomical Society}, vol.~367,
  no.~4, pp.~1766--1780, 2006.

\bibitem{thompson86}
A.~Thompson, J.~Moran, and G.~Swenson, eds., {\em Interferometry and Synthesis
  in Radio astronomy}.
\newblock John Wiley and Sons, 1986.

\bibitem{taylor99}
G.~Taylor, C.~Carilli, and R.~Perley, {\em Synthesis Imaging in
  Radio-Astronomy}.
\newblock Astronomical Society of the Pacific, 1999.

\bibitem{leshem2000b}
A.~Leshem, A.~van~der Veen, and A.~J. Boonstra, ``Multichannel interference
  mitigation techniques in radio-astronomy,'' {\em The Astrophysical Journal
  Supplements}, pp.~355--373, November 2000.

\bibitem{vandertol2009}
S.~{van der Tol}, {\em Bayesian Estimation for Ionospheric Calibration in Radio
  Astronomy}.
\newblock PhD thesis, Delft University of Technology, 2009.

\bibitem{wijnholds2009}
R.~N. S.~Wijnholds, S. van der~Tol and A.-J. van~der Veen, ``Calibration
  challenges for next generation of radio telescopes,'' {\em IEEE Signal
  Processing magazine}, Submitted 2009.

\bibitem{cornwell2008a}
T.~Cornwell, ``Multiscale {CLEAN} deconvolution of radio synthesis images,''
  {\em Selected Topics in Signal Processing, IEEE Journal of}, vol.~2,
  pp.~793--801, Oct. 2008.

\bibitem{Clark1980}
B.~G. Clark, ``An efficient implementation of the algorithm "clean",'' {\em
  Astronomy and Astrophysics}, vol.~89, pp.~377--378, 1980.

\bibitem{Schwab1984}
F.~R. Schwab, ``Relaxing the isoplanatism assumption in self-calibration;
  applications to low-frequency radio interfetomerty,'' {\em The Astronomical
  Journal}, vol.~89, pp.~1076--1081, July 1984.

\bibitem{cotton2008}
W.~D. {Cotton} and J.~M. {Uson}, ``{Pixelization and dynamic range in radio
  interferometry},'' {\em Astronomy and Astrophysics}, vol.~490, pp.~455--460,
  Oct. 2008.

\bibitem{voronkov2004}
M.~A. {Voronkov} and M.~H. {Wieringa}, ``"the cotton-schwab clean at ultra-high
  dynamic range",'' {\em Experimental Astronomy}, vol.~18, pp.~13--29, apr
  2004.

\bibitem{Cornwell2008}
T.~Cornwell, K.~Golap, and S.~Bhatnagar, ``The non-coplanar baselines effect in
  radio interferometry: The w-projetion algorithm,'' {\em IEEE Journal of
  Selected Topics in Signal Processing}, vol.~2, pp.~647--657, October 2008.

\bibitem{Frater1980}
R.~H. Frater and I.~S. Docherty, ``On the reduction of three dimensional
  interferometer measurements,'' {\em Astrononomy and Astrophysics}, vol.~84,
  pp.~75--77, Apr. 1980.

\bibitem{bhatnagar04}
S.~Bhatnager and T.~Cornwell, ``Adaptive scale sensitive deconvolution of
  interferometric images {I}. {A}daptive scale pixel (asp) decomposition,''
  {\em Astronomy and Astrophysics}, vol.~426, pp.~747--754, 2004.

\bibitem{jaynes57}
E.~Jaynes, ``Information theory and statistical mechanics,'' {\em Physics
  Review}, vol.~106, pp.~620--630, May 1957.

\bibitem{jaynes82}
E.~T. Jaynes, ``On the rational of maximum-entropy methods,'' {\em Proceedings
  of the {IEEE}}, vol.~70, pp.~939--952, sep 1982.

\bibitem{roberts84}
J.~Roberts, ed., {\em Indirect imaging}.
\newblock Cambridge university press, 1984.

\bibitem{narayan86}
R.~Narayan and R.~Nityananda, ``Maximum entropy image restoration in
  astronomy,'' {\em Annual review of of Astronomy and Astrophysics}, vol.~24,
  pp.~127--170, 1986.

\bibitem{sault90}
R.~Sault, ``A modification of the {C}ornwell and {E}vans maximum entropy
  algorithm,'' {\em The Astrophysical Journal}, vol.~354, pp.~L61--63, 1990.

\bibitem{skilling84}
J.~Skilling and R.~Bryan", ``maximum entropy image restoration algorithm,''
  {\em Monthly Notices of the Royal Astronomical Society}, vol.~211,
  pp.~111--124, 1984.

\bibitem{maisinger2004}
K.~{Maisinger}, M.~P. {Hobson}, and A.~N. {Lasenby}, ``{Maximum-entropy image
  reconstruction using wavelets},'' {\em Monthly Notices of the Royal
  Astronomical Society}, vol.~347, pp.~339--354, Jan. 2004.

\bibitem{Marsh1987}
K.~Marsh and J.~Richardson, ``The objective function implicit in the {CLEAN}
  algorithm,'' {\em Astronomy and Astrophysics (ISSN 0004-6361)}, vol.~182,
  pp.~174--178, Aug 1987.

\bibitem{Dobson1996}
D.~Dobson and F.~Santosa, ``Recovery of blocky images from noisy and blurred
  data,'' {\em SIAM Journal on Applied Mathematics}, vol.~56, no.~4,
  pp.~1181--1198, 1996.

\bibitem{Chen1998}
S.~Chen, D.~Donoho, and M.~Saunders, ``Atomic decomposition by basis pursuit,''
  {\em Siam J. Sci Comput}, vol.~20, no.~1, pp.~33--61, 1998.

\bibitem{Feuer2003}
A.~Feuer and A.~Nemirovski, ``On sparse representation in pairs of bases,''
  {\em IEEE trasactions on information theory}, vol.~49, pp.~1579--1581, June
  2003.

\bibitem{Elad}
M.~Elad and A.~Bruckstein, ``A generalized uncertainty principle and sparse
  representation in pairs of bases,'' {\em IEEE trasactions on information
  theory}.

\bibitem{Rudelson2006}
M.~Rudelson and R.~Vershynim, ``Sparse reconstruction by convex relaxation:
  Fourier and gaussian measurements,'' Mar 2006.

\bibitem{leshem2004}
A.~van~der Veen, A.~Leshem, and A.~Boonstra, ``Array signal processing in
  radio-astronomy,'' {\em Experimental Astronomy}, vol.~17, pp.~231--249, June
  2004.

\bibitem{stoica03}
P.~Stoica, Z.~Wang, and J.~Li, ``Robust {C}apon beamforming,'' {\em IEEE Signal
  Processing Letters}, vol.~10, pp.~172--175, Jun. 2003.

\bibitem{boyd96}
L.~Vandenberghe and S.~Boyd, ``Semidefinite programming,'' {\em SIAM Review},
  vol.~38, pp.~49--95, Mar. 1996.

\bibitem{pearson84}
T.~Pearson and A.~Readhead, ``Image formation by self-calibration in radio
  astronomy,'' {\em Annual Rev. Astronomy and Astrophysics}, vol.~22,
  pp.~97--130, 1984.

\bibitem{li2004}
J.~Li, P.~Stoica, and Z.~Wang, ``Doubly constrained robust {C}apon
  beamformer,'' {\em IEEE Transactions on Signal Processing}, vol.~52,
  pp.~2407--2423, Sept. 2004.

\bibitem{rottgering1994}
H.~{Rottgering}, I.~{Snellen}, G.~{Miley}, J.~P. {de Jong}, R.~J. {Hanisch},
  and R.~{Perley}, ``{{VLA} observations of the rich X-ray cluster Abell
  2256},'' {\em Astrophysical Journal}, vol.~436, pp.~654--668, Dec. 1994.

\bibitem{clarke2006}
T.~Clarke and T.~Ensslin, ``Deep 1.4 {GHz} very large array observations of the
  radio halo and relic in {A}bell 2256,'' {\em The Astronomical Journal},
  vol.~131, pp.~2900--2912, June 2006.

\bibitem{bridle76}
A.~H. {Bridle} and E.~B. {Fomalont}, ``{Complex radio emission from the X-ray
  cluster Abell 2256},'' {\em Astronomy and Astrophysics}, vol.~52,
  pp.~107--113, Oct. 1976.

\end{thebibliography}

\end{document}